\documentclass[prb,aps,twocolumn,superscriptaddress]{revtex4-2}
\usepackage{graphicx}
\usepackage{wrapfig}
\usepackage{etoolbox}
\usepackage{xcolor}

\makeatletter
\appto\abstract{%
	\let\latexlist\list \rightskip=\leftskip
	\def\list{\edef\keeprightskip{\the\rightskip}\latexlist}%
	\patchcmd\latexlist{\ignorespaces}{\rightskip\keeprightskip\ignorespaces}{}{}%
}
\makeatother

\begin{document}
	
\title{Experimental set-up for thermal measurements at the nanoscale using an SThM probe with niobium nitride thermometer}
	
\author{R. Swami}
\affiliation{Institut N\'eel, CNRS, 25 avenue des Martyrs, 38042 Grenoble, France}
\affiliation{Universit\'e Grenoble Alpes, Institut N\'eel, 38042 Grenoble, France}

\author{G. Juli\'e}
\affiliation{Institut N\'eel, CNRS, 25 avenue des Martyrs, 38042 Grenoble, France}
\affiliation{Universit\'e Grenoble Alpes, Institut N\'eel, 38042 Grenoble, France}

\author{S. Le-Denmat}
\affiliation{Institut N\'eel, CNRS, 25 avenue des Martyrs, 38042 Grenoble, France}
\affiliation{Universit\'e Grenoble Alpes, Institut N\'eel, 38042 Grenoble, France}
	
\author{G. Pernot}
\affiliation{Universit\'e de Lorraine, CNRS, LEMTA, 54000 Nancy, France}
	
\author{D. Singhal}
\affiliation{Institut N\'eel, CNRS, 25 avenue des Martyrs, 38042 Grenoble, France}
\affiliation{Universit\'e Grenoble Alpes, Institut N\'eel, 38042 Grenoble, France}
	
\author{J. Paterson}
\affiliation{Institut N\'eel, CNRS, 25 avenue des Martyrs, 38042 Grenoble, France}
\affiliation{Universit\'e Grenoble Alpes, Institut N\'eel, 38042 Grenoble, France}

\author{J. Maire}
\affiliation{Institut N\'eel, CNRS, 25 avenue des Martyrs, 38042 Grenoble, France}
\affiliation{Universit\'e Grenoble Alpes, Institut N\'eel, 38042 Grenoble, France}
	
\author{J.F. Motte}
\affiliation{Institut N\'eel, CNRS, 25 avenue des Martyrs, 38042 Grenoble, France}
\affiliation{Universit\'e Grenoble Alpes, Institut N\'eel, 38042 Grenoble, France}
	
\author{N. Paillet}
\affiliation{Institut N\'eel, CNRS, 25 avenue des Martyrs, 38042 Grenoble, France}
\affiliation{Universit\'e Grenoble Alpes, Institut N\'eel, 38042 Grenoble, France}

\author{H. Guillou}
\affiliation{Institut N\'eel, CNRS, 25 avenue des Martyrs, 38042 Grenoble, France}
\affiliation{Universit\'e Grenoble Alpes, Institut N\'eel, 38042 Grenoble, France}
	
\author{S. Gom\`es}
\affiliation{CETHIL, CNRS, 9 Rue de la Physique, 69621 Villeurbanne, France}
	
\author{O. Bourgeois}
\affiliation{Institut N\'eel, CNRS, 25 avenue des Martyrs, 38042 Grenoble, France}
\affiliation{Universit\'e Grenoble Alpes, Institut N\'eel, 38042 Grenoble, France}
	
\email{olivier.bourgeois@neel.cnrs.fr}
	
\date{\today}
	
\begin{abstract}
	
Scanning Thermal Microscopy (SThM) has become an important measurement tool for characterizing the thermal properties of materials at the nanometer scale. This technique requires a SThM probe that combines an Atomic Force Microscopy (AFM) probe and a very sensitive resistive thermometry; the thermometer being located at the apex of the probe tip allows the mapping of temperature or thermal properties of nanostructured materials with very high spatial resolution. The high interest of the SThM technique in the field of thermal nanoscience currently suffers from a low temperature sensitivity despite its high spatial resolution. To address this challenge, we developed a high vacuum-based AFM system hosting a highly sensitive niobium nitride (NbN) SThM probe to demonstrate its unique performance. As a proof of concept, we utilized this custom-built system to carry out thermal measurements using the 3$\omega$ method. By measuring the $V_{3\omega}$ voltage on the NbN resistive thermometer in vacuum conditions, we were able to determine the SThM probe's thermal conductance and thermal time constant. The performance of the probe is demonstrated by doing thermal measurements in-contact with a sapphire sample.

\end{abstract}
	
	
\maketitle

\section{INTRODUCTION}
	
Mastering local temperature fields and thermo-physical phenomena in nanometric-scale devices and structures plays a crucial role in growth of new technologies involving thermal issues \cite{pop2010energy,bourgeois2016reduction,zardo2019manipulating}.
One of the most critical examples is the production of unwanted heat in electronic transistors due to transport of electrons, which reduces the lifetime or induces failures in integrated electronic circuits and devices \cite{moore2014emerging}. In addition, manipulating heat transport in nanostructures is of crucial interest, so as to utilize them for thermal energy conversion, storage technologies such as solid-state thermoelectric power nanogenerators \cite{davila2012monolithically,perez2014micropower,varpula2017thermoelectric,tainoff2019network} or for local refrigeration based on the Seebeck and Peltier effects \cite{singhal2019nanowire,harzheim2020direct}. 
These developments require measurements of heat flow or thermal conductivity at the nanometer scale. Concerning heat transport in nano-contact or atomic junctions, numerous debated issues have been raised recently that clearly require new tools for further investigations \cite{menges2016nanoscale,tavakoli2018,cui2019thermal,gehring2019efficient}. Therefore, precise thermal measurements from micrometer to nanometer scales are essential for fundamental understanding of phonon heat transport but also for the optimized development of novel microelectronic, optoelectronic and innovative thermal nanodevices.

Among many different existing thermal measurement techniques \cite{cahill2014,zhao2016measurement}, scanning thermal microscopy (SThM) is one of the best available tool to probe temperature and thermal properties of materials with high spatial resolution at very small length scales \cite{majumdar1999scanning,gomes2015scanning,zhang2020review}. The first development towards the thermal sensing was introduced by Williams and Wickramasinghe in 1986 \cite{williams1986scanning}. They designed a probe for mapping the topography images of conducting and non-conducting materials to resolve the shortcomings of the Scanning Tunneling Microscopy (STM) technique, which is limited to conducting materials. The thermal probe was fabricated by placing a thermocouple junction at the end of the conventional STM tip, and the technique was known as Scanning Thermal Profiler (STP) \cite{williams1986scanning}. In this technique, the temperature of the thermal probe works as an intermediate quantity for maintaining constant tip-sample distance, in a similar way to a STM, where the tunnelling current is used to regulate the height of the tip. Although the STP technique could not be used to make the thermal maps of the surface, this was the beginning of the Scanning Probe Microscopy (SPM)-based technique in thermal science. Afterwards, continuous efforts have been carried out in the field of thermal microscopy based on STM \cite{weaver1989optical}. However, the major disadvantage of this technique is that an STM-based system can only be applied to electrically conducting samples. Conversely, the Atomic Force Microscopy (AFM)  technique can be utilized on all kinds of solid materials such as metals, semi-metals, semiconductors and insulators. Therefore, AFM-based microscopy is preferred over STM for dielectrics. Since AFM is proven to be a versatile technique as various sensors at the AFM probe tip could be integrated, multiple aspects of localized measurement will be allowed, especially thermal ones \cite{zhou1998generic}.

A breakthrough in the field of thermal science took place in 1993 when Majumdar \textit{et al.} \cite{majumdar1993thermal} developed a thermal probe by replacing the conventional AFM probe with two metallic wires to form a thermocouple junction. This probe was adequate to map the surface topography and the thermal map of a metal-semiconductor field-effect transistor simultaneously. Notably, this AFM-based thermal microscope proposed by Majumdar \textit{et al.} \cite{majumdar1993thermal} could be accounted as a building block for the modern SThM setups. Its working principle is based on scanning the surface of a specific sample with a heat-sensitive AFM probe. A temperature difference between the probe and the sample creates a heat flow that induces a change in the probe temperature-dependent physical property. For SThM, several types of thermo-sensitive sensors have been developed based on temperature dependent physical properties such as thermoelectric voltage  \cite{majumdar1993thermal,shi2001design,kim2012ultra,nguyen2019}, electrical resistance \cite{pylkki1994scanning,lefevre20053,menges2016nanoscale,janus2018,bodzenta2020,pernot2021frequency}, thermal expansion \cite{nakabeppu1995scanning} and fluorescence \cite{samson2008ac}. 

Among various possible thermometers, electrical resistance-based thermal sensors are the most widely used for their convenience of fabrication, use and sensitivity. The use of such sensors on AFM cantilevers for efficient SThM measurements requires a high temperature coefficient of the sensor's electrical resistance (TCR), low thermal conductivity of the probe materials  and, in this work, electrically insulated support. This implies that it is preferable to use a cantilever with a high thermal resistivity material, such as SiN. Commonly used materials for thermal sensors include metallic thin films of platinum  \cite{hatakeyama2014wafer}, alloy of platinum-rhodium (Pt90/Rh10)  \cite{pylkki1994scanning,lefevre20053}, palladium  \cite{dobson2007new,puyoo2010thermal}, low doped silicon probes \cite{menges2016nanoscale} and alloy of platinum-carbon (Pt-C) \cite{rangelow2001thermal}. The commercially available probes are of three kinds: either the Wollaston wire probe, palladium probes (Pd probes) with a SiN cantilever or doped Si-probes (DS probes) \cite{lefevre2004probe,dobson2007new,menges2016nanoscale}. The Wollaston wire probe was the first commercialized probe; the cantilever is composed of a U-shaped Wollaston wire 75~$\mu$m in diameter and the tip, which is also the probe's thermoresitive sensor, it consists of a V-shaped platinum-rhodium (Pt90/Rh10) filament obtained by electrochemical etching of the Wollaston wire. The documented spatial resolution and TCR of Wollaston wire probes are about 0.5~$\mu$m and $1.7\times10^{-3}$K$^{-1}$ respectively \cite{lefevre2004probe}. Meanwhile, for Pd probes, the cantilever is made of 500~nm thick SiN with an integrated Pd thin film at the apex of the tip which acts as a resistive thermal sensor. It can provide thermal images with sub-100~nm spatial resolution and its TCR is reported to be about $1.2\times10^{-3}$K$^{-1}$ ~\cite{dobson2007new}. Concerning DS probes, the cantilever is micromachined in a U shape with high doping level and a low doped platform, which operates as the thermal sensor \cite{menges2016nanoscale}. The radius and height of the Si tip, which is supported on the low doped platform, are about 10~nm and 500~nm, allowing the mapping of materials with very high spatial resolution, but these probes suffer from a relatively low TCR of $1.2\times10^{-3}$K$^{-1}$ and low thermal resistance of the cantilever \cite{nelson2007measuring}.


Finally, through a compromise between thermal and spatial resolutions, available commercial thermal probes can reach high spatial resolution but exhibit a meagre thermal sensitivity. Such low thermal sensitivities compel the need of a high thermal exchange between the probe and the sample, \textit{i.e.} a large temperature difference is required, excluding measurements at low temperature differences (temperature oscillations smaller than 10~K). Indeed, as pointed out by Shi \textit{et al.} \cite{shi2001design}, efficient SThM probes have to be fabricated using low thermal conductivity materials, like SiN or Si$\rm O_{2}$, in order to minimize heat loss through the cantilever. Besides, newly developed SThM probes require innovative thermometry to reach drastic improvements in their thermal sensitivity \cite{swami2022electron} along with low-noise electronics for the acquisition of the thermal signal with high accuracy.

\begin{table*}
\begin{center}
\begin{tabular}{ | l | c |  c |  c | }

\hline
\hline\noalign{\smallskip}
 SThM probe type & DS \cite{menges2016nanoscale} & Pd probe \cite{dobson2007new} & NbN probe \cite{swami2022electron} \\
 
\hline\noalign{\smallskip}

Thermal sensor & doped Si & Pd & NbN \\

 \hline
 
Tip radius ($\mu$m) &  0.02 & 0.1 & 0.05/0.2  \\

Tip height ($\mu$m)  & 0.5   & 10 & 3 \\

Sping constant (Nm$^{-1}$) & 1 & 0.35 & 0.15   \\

Resistance (kOhm) & 0.5 & 0.4 & 5 to 20   \\

TCR ($\times 10^{-3}$ K$^{-1}$) & 1.2 & 1.2 & 4.5   \\

\hline
\hline
	\end{tabular}
		\caption{Comparisons of probe's parameters between DS, Pd and the NbN based probe of this work.}
\label{tableSiN}
\end{center}
\end{table*}

To meet these requirements, we developed in this work a highly sensitive electro-thermal measurement setup by combining a thermal probe equipped with a highly sensitive niobium nitride (NbN) thermometer in a high vacuum-based scanning thermal microscopy system. The SThM setup was custom-built by modifying an EnviroScope SPM scanning probe microscope system, which involved designing and fabricating an adapted probe holder and integrating low-noise electronics for the data acquisition for thermal measurements, as it will be described in the following. We will show the temperature calibration of the electrical resistance of the highly sensitive NbN sensor integrated on flat substrate and at the tip of the AFM probe. Finally, we employed the dynamic 3$\omega$-SThM mode under high vacuum conditions to determine the thermal properties of the probe \cite{lefevre20053,chirtoc2008,puyoo2010thermal,bodzenta2016,pernot2021frequency}. The measured thermal properties were compared with theoretically calculated values based on the probe's geometry and composition. Results show that SThM experiments allow to obtain thermal properties at the nanoscale, as it will be demonstrated on a sapphire sample.

\begin{figure*}[!ht]
	\centering
	\includegraphics[width=0.9\textwidth]{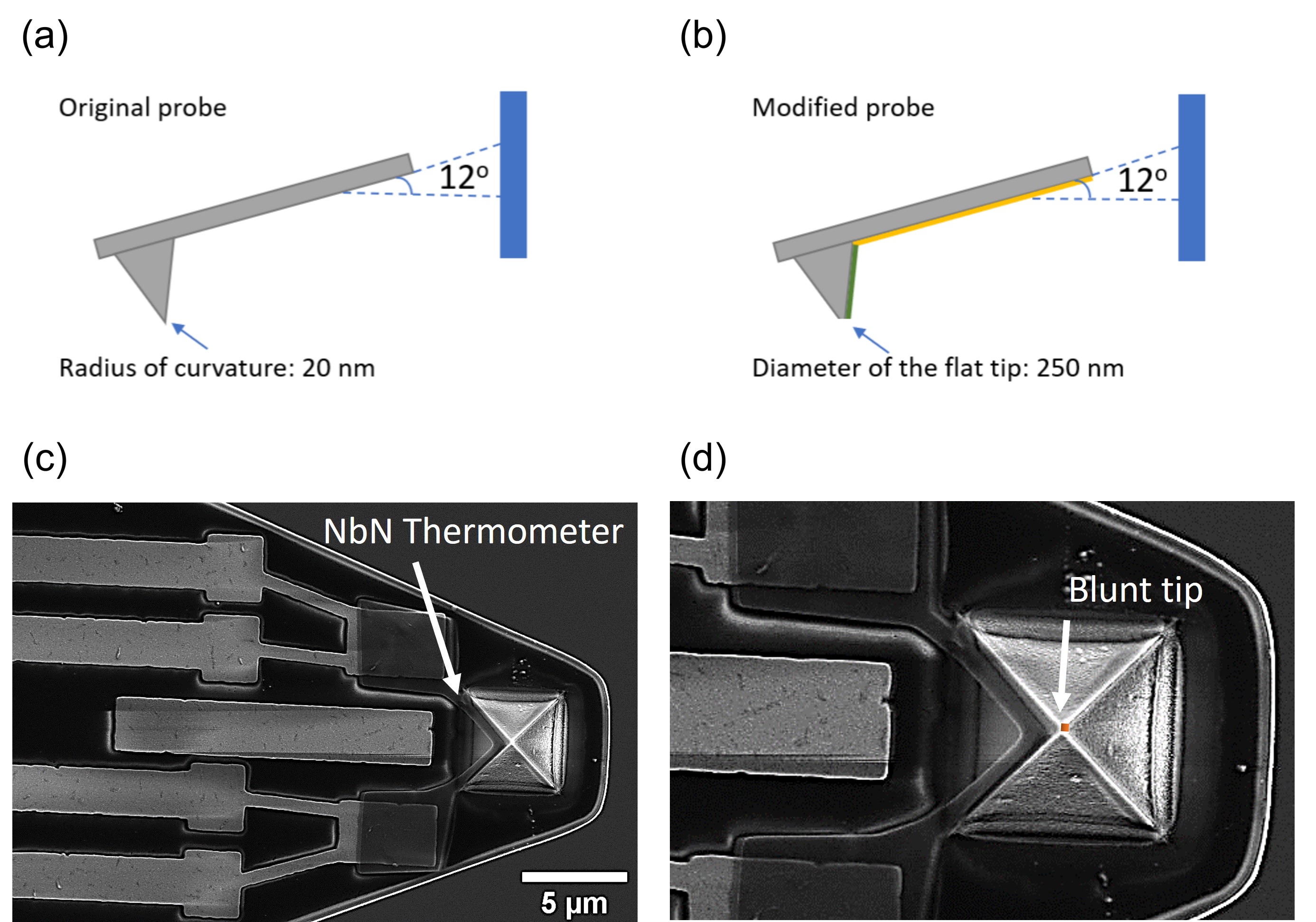}
	\caption{Schematic illustration of the original AFM probe (a), and (b) FIB modified SThM probe having an equivalent diameter of tip about 250~nm, the yellow line indicates the gold leads and the green line the NbN thermometer located on one face of a square based pyramid. (c) and (d) are the SEM images of the NbN-SThM probe with a blunt tip with a square flat apex of about 250~nm side. (c) Top-view of AFM probe with the triangular-shaped cantilever and tip. (d) Close-up SEM micrograph of the blunt tip with an integrated NbN thermometer near the apex of the tip.}
	\label{Fig:2}
\end{figure*}

\section{SThM experimental set-up}

The first purpose of this work is to adapt the EnviroScope SPM instrument for quantitative thermal measurements at the nanoscale. In order to achieve this, we had to integrate the thermal control unit in the AFM instrument. The AFM instrument has been modified into SThM setup by integrating a thermal probe and thermal control unit. The thermal control unit incorporates the following elements: (a) a specific probe holder to install the SThM probe, the holder being used to establish an electrical connection between the probe and the external electronics, (b) low noise electronics that consist of several instruments such as a lock-in amplifier, a voltage to current converter able to deliver nanoampere, a variable resistance box and a differential bridge adapted to the targeted resistance of the NbN thermometer in the range 5 to 20 kOhm. These different elements will be described in details in the following.

\subsection{Micro and nanofabrication of the NbN thermal probe}

Experiments will be carried out using NbN based SThM probes installed in the EnviroScope SPM instrument. The NbN probes are fabricated following a process that has been already described in a previous article \cite{swami2022electron} from a commercial SiN AFM probes OTR8-10 from Bruker \textregistered~having low thermal conductivity and electrical insulating properties. These probes have triangular-shaped cantilevers and pyramidal-shaped tips with a tip height of 2.5 to 3.5~$\mu$m. The length and width of these SiN cantilevers are 140~$\mu$m and 30~$\mu$m respectively, and 800~nm thick; the apex of the tip having a radius of curvature about 15~nm to 20~nm. The first characterizations of the cantilever indicate a resonance frequency of 22.33 kHz and a spring constant of 0.1544 Nm$^{-1}$. Since all the measurements will be performed under vacuum at a pressure of 5~$\times10^{-6}$~mbar, there will be no water meniscus and consequently the thermal exchange will be significantly smaller than measurements under air conditions (no heat transfer between the probe and the sample through the water meniscus that can condense at the contact and through the air) \cite{AssyAPL2015,AssyNanotech2015}; this is an advantage to get more reproducible thermal measurements. However, the drawback is that the effective contact surface that will contribute to the thermal transport between the tip and the surface is {\it de-facto} reduced, the overall experiment being more sensitive to roughness and sample surface state \cite{gotsmann,gomes2015scanning,zhang2020review,guen}. We decided then to work with a blunt tip to increase the exchange surface and hence the heat flow in a controllable way. Cutting part of the apex of the tip modifies the radius of curvature with an increased exchange surface, a square flat apex of about 250~nm side, as shown in Figure~\ref{Fig:2}. To proceed with the fabrication of the blunt tip, one of the most critical parameters to take into consideration is the angle of the cutting plane. In our AFM equipment, the installed probe has a tilted angle of 12$^{\circ}$ relative to the horizontal plane. In this respect, we considered this angle to cut properly the tip so that when the probe touches the sample, the size and shape of the contact of the tip and the sample surface are parallel, the contact being then equivalent to those of the tip apex. 

To cut the SThM tip, we used the dual-beam platforms named as LEO-1530 field emission SEM system, combining a Scanning Electron Microscope (SEM) and a Focus Ion Beam (FIB) column. After having performed that step, we then proceeded with the optimized fabrication process to integrate the 70~nm thick NbN thin thermometric film at the apex of the AFM tip along with electrical 50~nm thick gold leads on the cantilever in a four-wire configuration as already described in a previous publication \cite{swami2022electron}. Figure~\ref{Fig:2} shows the SEM-FIB images of the SThM probe having a blunt tip with a mean diameter of $\sim$250~nm. The contact surface area of the probe has been increased potentially by one order of magnitude to lower the thermal interface resistance between the tip and sample during the thermal measurements under vacuum. However, it is worth noticing that as this procedure is enhancing the heat flux from tip to sample, the spatial resolution of the thermal measurement is strongly degraded.

\begin{figure}[!ht]
	\includegraphics[width=0.5\textwidth]{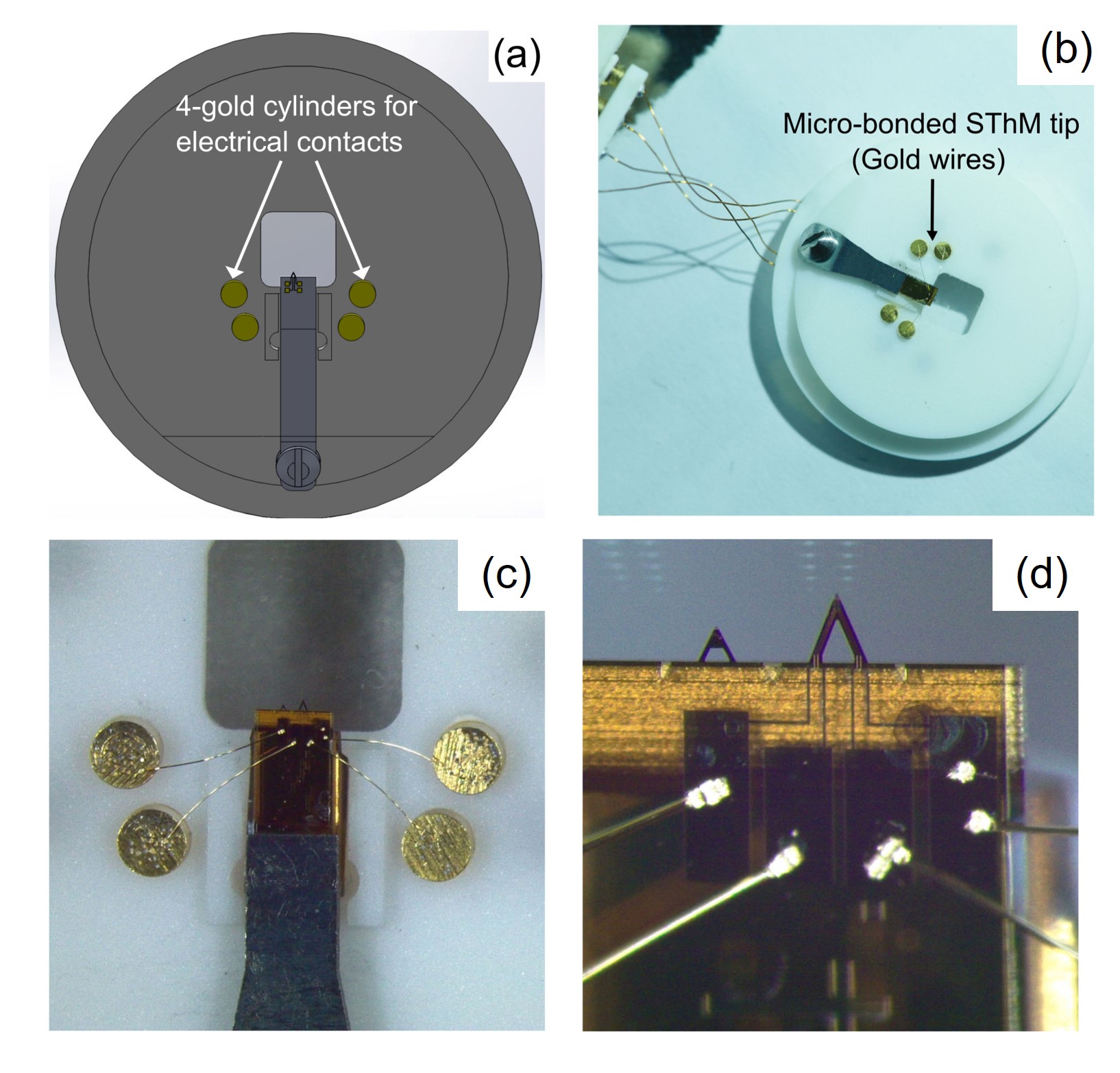}
	\caption{The geometry of a new SThM cantilever holder used for thermal measurements. (a) Top view consists of four gold-plated cylinders to establish the electrical contacts. (b) Photographs of the SThM probe installed on the probe holder with the Au wire bonding. (c) and (d) represent the close-up optical images of the wire bonded gold pads on the SThM chip.}
	\label{Fig:3}
\end{figure}

\subsection{Thermal Probe Holder}

One of the main modifications to the microscope was to design and manufacture a novel thermal cantilever holder capable of performing resistive thermometry using four-wire electrical measurements. The most common requirements of the thermal cantilever holder are lightweight, easy installation and exchange of thermal probe, and compatibility with laser alignment. Figures~\ref{Fig:3} (a) and (b) show the geometry of the probe holder constructed in SolidWorks\textregistered. The main body of the cantilever holder is made of Macor\textregistered~~ceramic-based glass with low thermal conductivity, electrically insulating, strong, rigid, and has zero porosity. This ceramic based-glass is well suited for measurements at ambient conditions as well as high-vacuum SThM experiments. On this holder, the thermal probe is installed using a metal clamp with a tightening screw and tilted 12$^{\circ}$ relative to the horizontal plane concerning the longest dimension. Since the SThM experiments will be performed using a resistive thermometer in a four-wire configuration, we have integrated four gold-plated cylinders in the holder for establishing the electrical connections by microbonding. In addition, a dither piezo is integrated on the backside of the holder for operating the thermal cantilever in dynamic mode if needed. Figures~\ref{Fig:3} (c) and (d) illustrate the photograph of an SThM probe installed on the probe holder where Au wires were used to microbond the probe to gold plated cylinders. Figures~\ref{Fig:3} (e) and (f) show the close-up optical images of the wire-bonded gold pads on the SThM chip.

\subsection{Low-noise electronics for data acquisition}

The second important task in this work was to set-up the proper measurement chain to characterize the thermometer and perform thermal measurements using the 3$\omega$ method \cite{cahill1990thermal,lefevre20053,bodzenta2020}. The electrical resistance of thermometer versus temperature will be obtained by a measurement based on a dynamic method using an AC current, the AC voltage being measured by a lock-in amplifier equipment. The thermal properties of the system are obtained by measuring the 3$\omega$ voltage, as it will be explained later, again requiring lock-in amplifier-based measurement chain.

In order to convert the AC voltage generated by the lock-in into a measuring current, we used an instrument developed at in-house that allows to convert both DC and AC voltages into DC and AC currents. This voltage to current converter, built to have a very low thermal drift of about 1~ppm/$^{\circ}$C, can provide very stable  electrical currents from 1~nA to 1~mA.

\begin{figure}[!ht]
	\centering
	\includegraphics[width=0.5\textwidth]{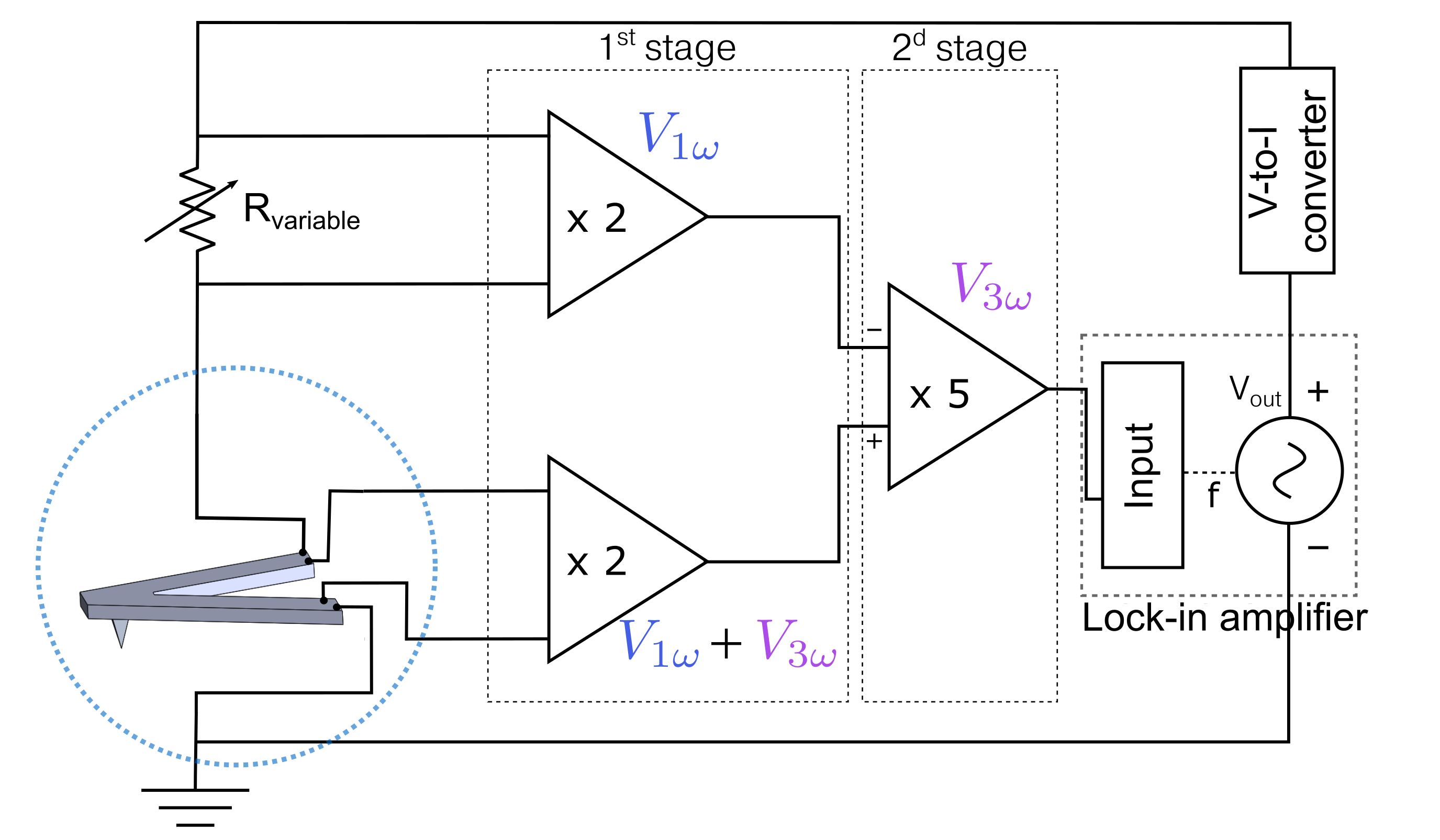}
	\caption{Schematic of the differential bridge used in the SThM experimental setup for 3$\omega$ experiments. The gain of the first stage of the amplifier is two and the gain of the second stage is five. The variable resistance is used to compensate the $V_{1\omega}$ coming from the SThM NbN resistance. V-to-I refers to the voltage to current converter.}
	\label{Fig:4}
\end{figure}

To accurately measure the electrical signal and improve the signal-to-noise ratio, various measurement bridges can be considered. The classical Wheatstone bridge is the most widely used for SThM measurements. It requires two known resistances with an additional variable resistance and a resistive thermal probe. The basic principle of using the Wheatstone bridge is based on balancing the bridge. For that, the resistance of the thermal probe is balanced by tuning the variable resistance. During the experiment, a slight variation in the SThM probe temperature can be detected with a high sensitivity. The disadvantage of this setup is that it cannot be utilized in a four-wire configuration \cite{sandell2020thermoreflectance}. Notably, four-wire geometry (like for the NbN probe) provides more localized measurements of the sensor electrical resistance since it does not include the electrical resistance of the electrical leads, the contact resistance and external measuring cables.

We then decided to use a homemade differential bridge setup adapted to two-terminal or four-terminal thermal probes built from three INA103 instrumentation amplifiers. Figure~\ref{Fig:4} illustrates the schematic of a simplified SThM setup with a current source, variable electrical resistance, lock-in amplifier, and the differential bridge that contains two amplification stages. When the differential bridge is balanced (the variable electrical resistance and the one of the thermal probe are equal), the output voltage at the first harmonic of the differential bridge is greatly reduced. Assuming that the variable resistance has a negligible temperature coefficient of resistance, the thermal signal will be obtained from the third harmonic of the output voltage of the differential bridge, as it will be explained later (see section III B). The amplification (gain~$\times$10) of the differential bridge has been chosen to ease the low noise detection with the lock-in amplifier.

\section{Thermal probe: calibration and characterization}

\subsection{Calibration of the NbN thermometers}

In this work, the NbN thin film used as a resistive thermometry sensor is the cornerstone of the SThM probe.
The NbN thin films are deposited by magnetron sputtering from a pure Nb target (99.95~\%) in a gas mixture composed of Ar (3~\%) and N (97~\%) \cite{swami2022electron,bourgeois2006liquid,nguyen2019niobium}. The thickness is approximately 70~nm for a targeted electrical resistivity at room temperature of approximately $7\times10^{-5}$~$\Omega$m. The calibration of the thermometer is done by measuring the NbN electrical resistance $R$ as a function of the temperature $T$. During the calibration procedure of $R$ versus $T$, the temperature of the NbN thin film is measured by a home-made temperature controller (TRMC2) using a calibrated platinum thermometer as a reference; the temperature being stable and regulated with a sensitivity better than 5~mK at 300~K measured with a time constant of 100~ms.

\begin{figure}[!ht]
	\includegraphics[width=0.5\textwidth]{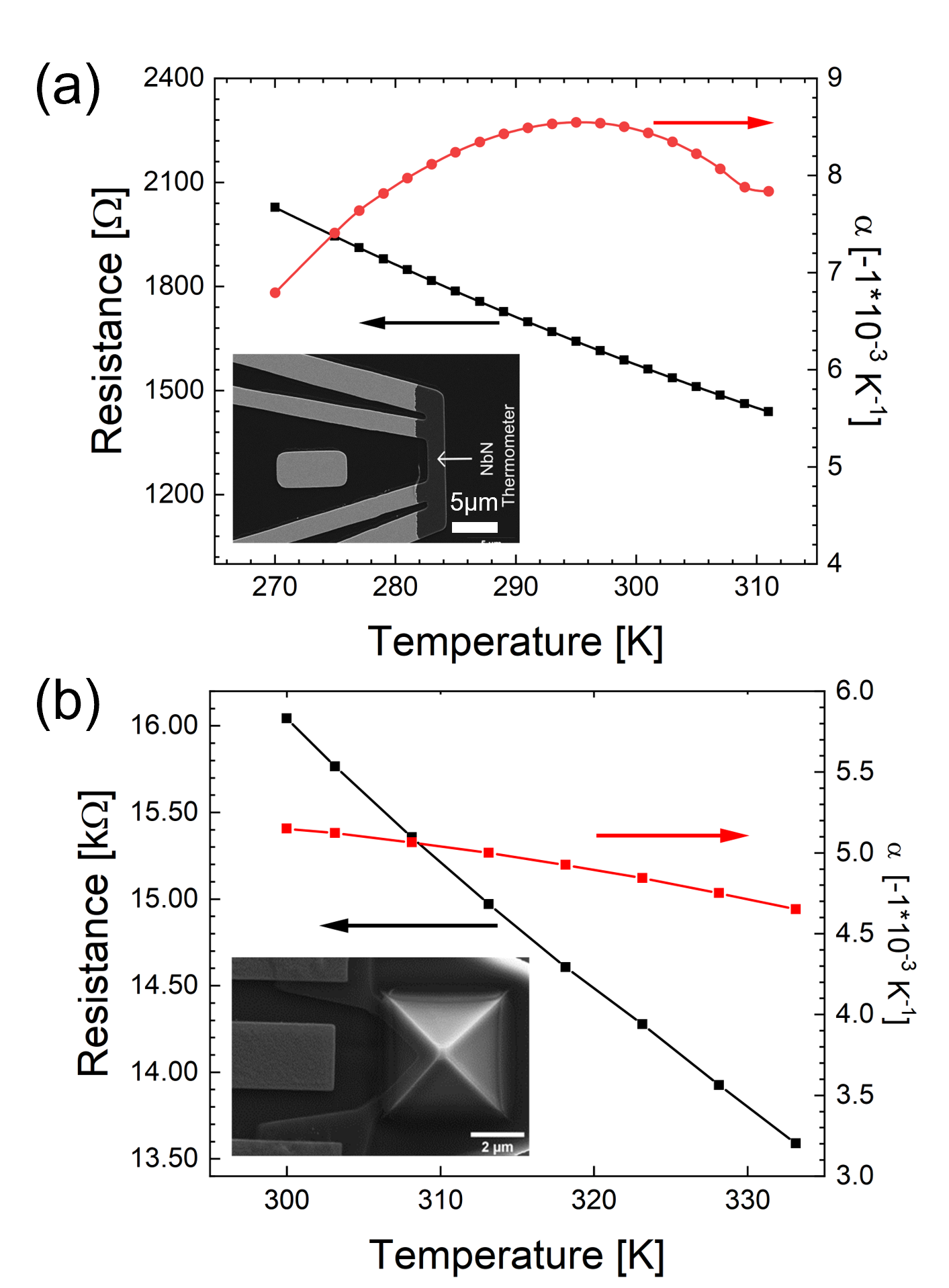}
	\caption{The electrical response and temperature coefficient of resistance of the NbN thermometers are illustrated as a function of temperature. (a) NbN thin film of thickness 70~nm was deposited on the flat Si/SiN substrate. The electrical resistance (black squares) and TCR (red squares) of NbN thin film are plotted as a function of temperature. The TCR was obtained to be around  $|8\times10^{-3}|$ K$^{-1}$ near room temperature. (b) The black squares represent the experimentally obtained values of electrical resistance of the NbN thermometer on the SThM probe. The red squares represent the TCR of the NbN thermometer of the SThM probe showing a value of TCR to be around $|4.5\times10^{-3}|$ K$^{-1}$ near room temperature. All lines are guides to the eyes.}
	\label{Fig:5}
\end{figure}

The performance of the NbN thermometer on the pyramid of the AFM probe is evaluated compared to NbN thermometers prepared on a regular Si/SiN flat surface. We want to check that depositing the NbN materials on the non-planar pyramid of the tip does not change significantly the material properties.

The electrical characterizations of integrated NbN resistance on a flat SiN substrate and AFM probe are conducted using a four-probe measurement setup by applying current into the outer leads and measuring the voltage drop from the inner leads in a furnace regulated in temperature with a resolution of 10~mK. The range of amplitudes of the applied current $I_{1\omega}=I_0 cos(\omega t)$ is chosen from 10~nA to 1~$\mu$A that leads to a power dissipation in the sensor of $P_{2\omega} (\omega)=R_{0}I_{1\omega}^{2}$ from few tens of picowatts to few nanowatts preventing any sensor overheating during the calibration procedure. The voltage drop is constantly monitored using a lock-in-amplifier; a linear I-V behavior (not shown) is observed in that current range, that indicates a negligible Joule heating inside the probe. 

Figure~\ref{Fig:5} (a) shows the calibration curve of the NbN thermometer deposited on a regular Si/SiN flat substrate while Figure~\ref{Fig:5} (b) shows the calibration curve of the NbN thermometer integrated at the apex of the AFM probe; in insets the corresponding SEM image of each sample. The electrical resistance of the NbN thermometer is plotted as a function of temperature, where black squares are the experimentally measured values of electrical resistance. The red squares represent the absolute value of TCR $|\alpha|$ defined as $\alpha=1/R\times dR/dT$ of the NbN thermometer on the flat surface, which is around $\alpha=-8\times10^{-3}$~K$^{-1}$ near room temperature, consistent with previously published results \cite{bourgeois2006liquid}. One should note that the negative TCR of the NbN thin film is a signature of a Mott-Anderson insulator materials, {\it i.e.} a type of materials that undergoes a transition to the insulating state as the temperature is lowered \cite{bourgeois2006liquid,nguyen2019niobium}.

Figure~\ref{Fig:5} (b) depicts the temperature-dependent electrical resistance (black squares) and TCR (red squares) of the SThM probe. We measured the change in the electrical resistance from 16~k$\Omega$ up to 14~k$\Omega$ in the temperature range of 305~K to 335~K, leading to an extracted absolute value of TCR of $|\alpha|=5\times10^{-3}$ K$^{-1}$ at room temperature; this thermometer resistance is much larger than the resistance of the gold leads (200~Ohm, measured in two wire geometry). In the following, the power dissipated in the gold leads will then be neglected. We can note that the measured electrical resistance and TCR of the NbN thermometer between the SThM probe and flat surface are different. This change of TCR could have various origins: variation of thickness or nitrogen contents in the NbN film. The observed decrease can be preferentially attributed to a higher amount of nitrogen contents in the NbN film on the pyramid leading to higher electrical resistance and, consequently, a lower TCR \cite{bourgeois2006liquid}. In the case of the SThM probe, a probable reason for the excess amount of nitrogen in the NbN film is potentially due to overheating at the AFM tip (thermally isolated compared to flat Si/SiN thick substrate) during the sputtering process.

\subsection{Experimental conditions for thermal measurements by 3$\omega$ method using the NbN SThM probe}

To qualify the potentialities of these NbN based SThM probes, we have performed thermal measurements, first, on the bare probe and then in contact with a sapphire sample of mean roughness less than 1~nm. The thermal measurements have been done by the 3$\omega$ method vastly used for the measurement of thermal conductance either on bulk substrate \cite{cahill1990thermal}, on membrane \cite{ftouni2015thermal} or on suspended nanowires \cite{heron2009mesoscopic}.

\begin{figure}[!ht]
	\includegraphics[width=0.4\textwidth]{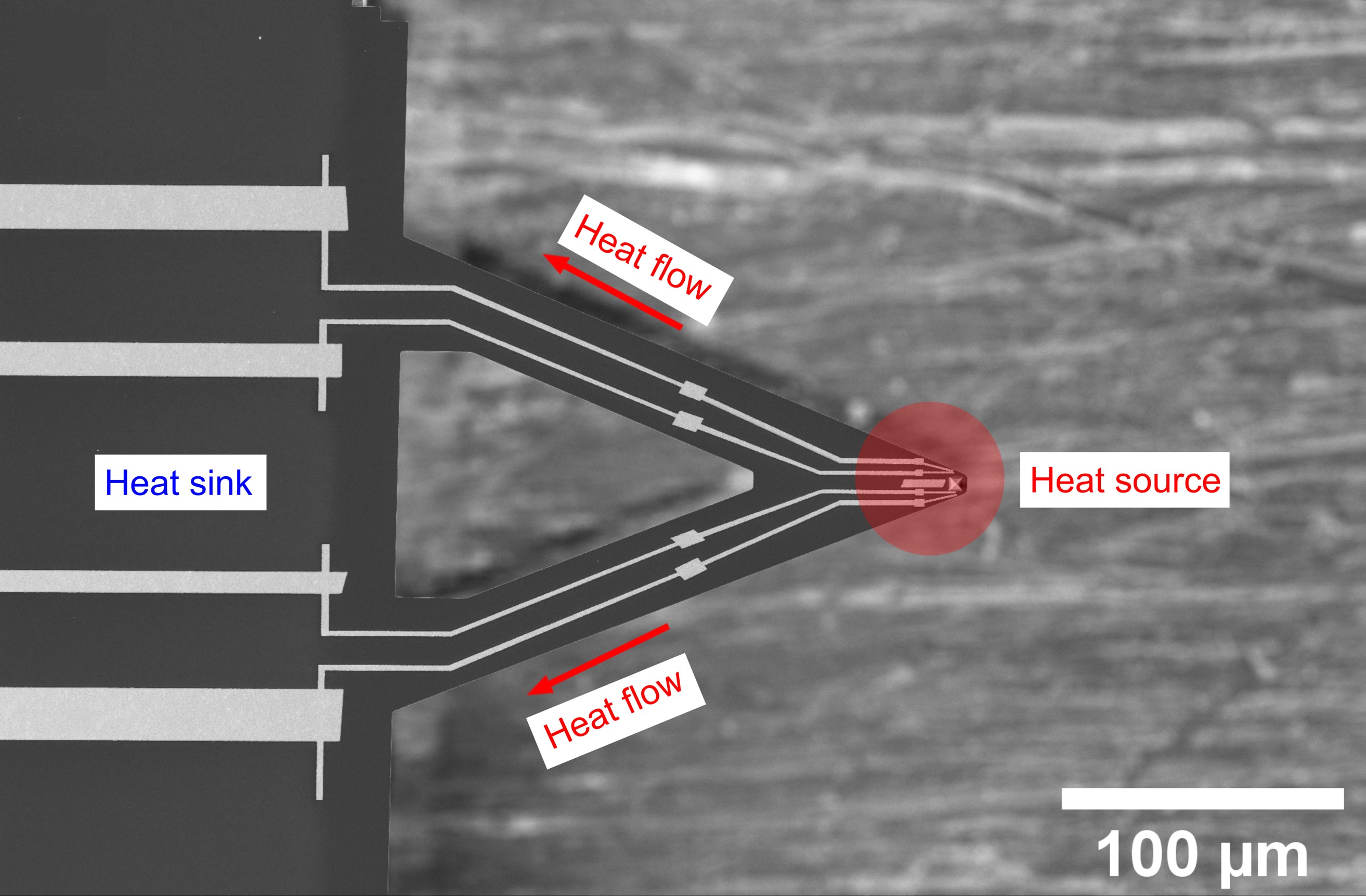}
	\caption{ SEM image of the SThM probe used in this study. The red circle indicates the location of the heat source, the cantilever as a thermal link and the chip body as a heat sink. The heat is transported from the hot part of the SThM probe (heat source) to the chip body (heat sink) in vacuum conditions through the arms of the cantilever.}
	\label{Fig:6}
\end{figure}

The principle of the 3$\omega$ method relies on employing a one-dimensional thin metal film deposited on a sample that acts both as a heater and thermometer to create a heat source as depicted in Figures~\ref{Fig:6}  through an SEM picture and a simple thermal model. In 1999, Fiege \textit{et al.} \cite{fiege1999} combined the SThM with 3$\omega$ method to perform quantitative thermal measurements with higher accuracy, a method further developed since then \cite{lefevre20053,puyoo2010thermal,chirtoc2008,bodzenta2020,pernot2021frequency}. The basic idea of this technique is the same as classical 3$\omega$: an alternating current of angular frequency $\omega$ is applied through the NbN metal film that results in heating due to the Joule effect. The heat generated in the thermometer film at the apex of the probe (see Figure~\ref{Fig:6}) induces temperature oscillations at angular frequency 2$\omega$ of the pyramid of the SThM probe. Therefore, temperature oscillation leads to a periodic change in the electrical resistance of the metal film at angular frequency 2$\omega$. Since the thermometer is biased by a current at angular frequency $\omega$, consequently, a voltage at angular frequency 3$\omega$ develops across the metallic line proportional to the temperature oscillation at 2$\omega$. From this temperature oscillation, the thermal properties of the system beneath the thermometric line are extracted.

Let us mention the important equations proper to this active 3$\omega$-SThM mode where the tip is self-heated by an AC current that results in significant Joule heating. As discussed above, since the current is given by $ I_{1\omega}=I_0 cos(\omega t)$ then the electrical power $P_{2\omega}(\omega)$ dissipated in the NbN thermometer resistance $R_{0}$ of the SThM probe reads as: 
\begin{equation}
	P_{2\omega}(\omega)=\frac{R_{0}I_{0}^{2}}{2} (1+cos(2\omega t)).
\label{power1}
\end{equation}

In the following, 
\begin{equation}
	P_{2\omega}=\frac{R_{0}I_{0}^{2}}{2}
	\label{powerampl}
\end{equation}
will refer to the amplitude of the power dissipated in the NbN thermometer. 

When the tip is not in contact with the sample, then the temperature oscillation in the probe is related to the heat flux and thermal conductance of the probe as:
\begin{equation}
    Q_{2\omega}^{probe} = G_{probe} T_{2\omega-oc}
	\label{equation2}
\end{equation}
where $T_{2\omega-oc}$ is the amplitude of the temperature oscillation of the SThM probe where $oc$ stands for out-of-contact. Then, the thermal conductance of the SThM probe can be determined by equating the heat flux through the cantilever ($Q_{2\omega}^{probe}$) to the electrical power $P_{2\omega}$ (at low enough frequency $f<5$~Hz) that is required to achieve the  temperature oscillation amplitude $ T_{2\omega-oc}$ of the NbN SThM probe above the room temperature:
$$ P_{2\omega} = G_{probe} \times T_{2\omega-oc} $$ 

\begin{equation}
	G_{probe} = \frac{ R_{0}I_{0}^{2}}{2 T_{2\omega-o.c}}
	\label{equation3}
\end{equation}
It is worth noticing that "low enough frequency" defines the frequency range where only a small phase shift (below 10~degrees) exists between the excitation current $ I_{1\omega}$ and the voltage response at 3$\omega$.

In SThM studies, the knowledge of the probe’s thermal conductance ($G_{probe}$) is critical to build the thermal model adapted to reproduce the measurements in order to identify sample temperature or thermal properties.
We can determine $G_{probe}$ by performing the experiment in vacuum conditions when the probe is in and out-of-contact configurations. Another important parameter is the thermal time constant of the probe defined as the time required to reach $1/e$ of the initial state after a step function change of temperature measured by the NbN thermometer. The thermal time constant governs the cut-off frequency of the probe as a thermal low-pass filter. We will measure and calculate this constant in the following using 3$\omega$-SThM measurement.

\begin{figure}[!ht]
	\includegraphics[width=0.5\textwidth]{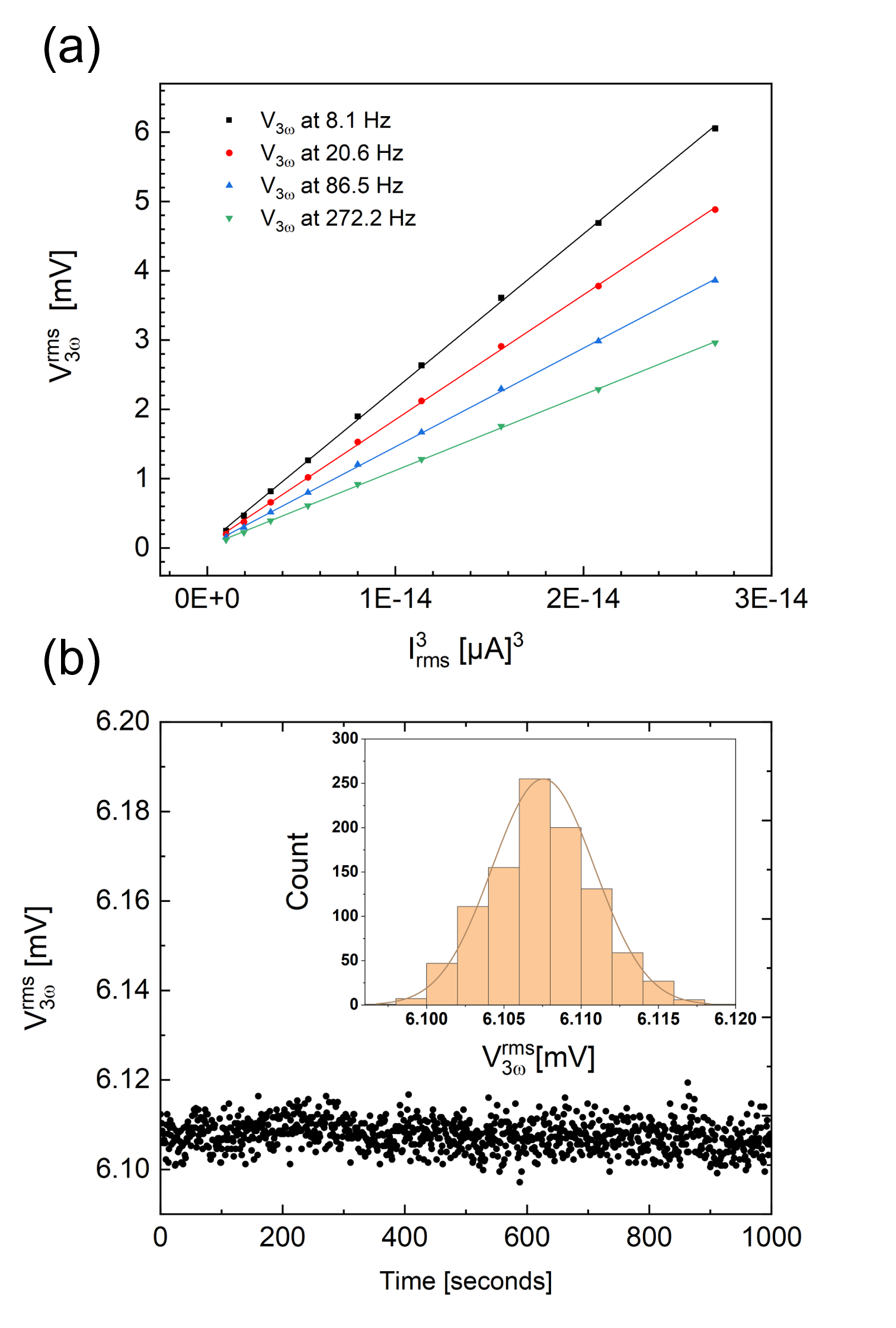}
	\caption{(a) Measurements of the linear behaviour of $V^{rms}_{3\omega}$ and $I_{rms}^{3}$ for different particular frequencies. (b) Measurement of the $V^{rms}_{3\omega}$ voltage noise in the SThM probe inside the scanning thermal microscope setup at a base pressure of 5~$\times10^{-6}$~mbar. Solid circles (black) are the $V^{rms}_{3\omega}$ voltage response of the thermal tip at 293~K as a function of time, measured using SR-830 lock-in amplifier. The inset represents the histogram (orange) of the thermal probe corresponding to its voltage response with a Gaussian fit (solid black line).}
	\label{Fig:7}
\end{figure}

 The experiment is performed in high-vacuum conditions ($3\times 10^{-6}$~mbar) when the probe is far from the sample. A predetermined AC current is supplied through the probe that results in Joule heating as illustrated in Figure~\ref{Fig:4}. We measured the $V^{rms}_{3 \omega}$ voltage amplitude response for various angular frequencies $\omega$, rms refers to the root mean square value. In order to confirm the reliability of the experiment, $V^{rms}_{3 \omega}$ signal should be linear with applied heating AC current $I_{rms}^3$ for a fixed frequency whatever the geometry of the transducers  \cite{cahill1990thermal,heron2009mesoscopic,ftouni2015thermal}. This has been checked, as shown in Figure~\ref{Fig:7} (a), where a linear variation of $V^{rms}_{3 \omega}$ is observed as a function of $I_{rms}^3$ at room temperature. The $V^{rms}_{3 \omega}$ voltage amplitude variation has been recorded at various electrical frequencies while keeping the heating rms AC current at a value of 30~$\mu$A. The noise on the ${3 \omega}$ voltage has also been measured as presented in Figure~\ref{Fig:7} (b) showing a standard deviation of $2\times 10^{-6}$ V/$\sqrt{\rm Hz}$. This noise is far above the expected Johnson noise ($\approx 16$~nV/$\sqrt{\rm Hz}$) for a resistance of 16~kOhm at room temperature indicating the presence of other noise sources like the vibrations of the experimental set-up, electrical contact resistance between the thermometer and the leads that need to be further investigated.

\subsection{Measurement of the thermal properties of the cantilever}

Now, by measuring the variation of $V_{3 \omega}$ as a function of the angular frequency, $\omega$ we will be able to extract an experimental thermal characteristic time of the SThM probe and compare it to the calculated one. Figure~\ref{Fig:8} shows the $V^{rms}_{3\omega}$ voltage response of the probe as a function of electrical frequency in the absence of a laser beam on the backside of the SThM cantilever. 

We used the experimental results presented in Figure~\ref{Fig:8} (a) obtained under vacuum conditions for determining the thermal conductance of the probe when the AFM laser is off. The electrical resistance $R_0$ of the thermometer of this probe was measured to be 16~kOhm at room temperature with a calibrated TCR value of 4.5$\times$ 10$^{-3}$K$^{-1}$. The  amplitude of the electrical power dissipated in the NbN thermometer was about 14.4~$\mu$W as calculated using Eq.~\ref{powerampl} that leads to increase the temperature at the apex of the tip. In the following equation, we take the low frequency limit of the temperature oscillation:
\begin{equation}
T_{2\omega}=\frac{2V^{rms}_{3\omega}}{R_0 I_{rms} \alpha} 
 \label{qprobe1}
\end{equation}
noting that the signal $T_{2\omega}$ barely changes with frequency up to 5~Hz.
We used Eq.~\ref{qprobe1} to determine the amplitude of the temperature oscillation of the probe $T_{2\omega}$ estimated to be about 5.9~K.  
When the probe is not in contact with the sample under vacuum conditions, then, the electric power ($P_{2\omega}$) is equivalent to the heat that flows from the NbN thermometer along the cantilever to the heat bath (the chip):
\begin{equation}
 P_{2\omega}= Q_{probe} = G_{probe} T_{2\omega} 
 \label{qprobe2}
\end{equation}
Using Eq.~\ref{qprobe2}, knowing the applied power and the measured $T_{2\omega}$, the thermal conductance of the probe is approximately 2.4$\times 10^{-6}$WK$^{-1}$. 

The experimental thermal time constant of the probe can be defined as $\tau_{th} = (2\pi f_{cutoff})^{-1}$, where $f_{cutoff}$ is the cut-off frequency at which the V$_{3 \omega}$ voltage is attenuated by a factor of $\sqrt{2}$ of the maximum value \cite{kim2012ultra}. The experimentally obtained value of the cut-off frequency is about 60~Hz (see Figure~\ref{Fig:8} (a)), and the measured thermal time constant of the SThM probe is about 3~ms, a value in relatively good agreement with the order of magnitude of few milliseconds found from rough calculations, as we will show below. 

\subsection{Calculation of the expected thermal conductance and thermal time constant}

Here, we will calculate the expected probe’s thermal conductance and thermal time constant using the geometry, dimensions, and thermal conductivity of the various materials from which the probe is made. 
For simplicity, we first divided the SiN cantilever into two rectangular plates, having length, width, and thickness of 200~$\mu$m, 40~$\mu$m, and 0.8~$\mu$m respectively. Second, we used the literature value of the thermal conductivity ($\kappa$) and specific heat ($C_p$) of the SiN, which are approximately 3~Wm$^{-1}$K$^{-1}$ and 0.8~Jg$^{-1}$K$^{-1}$ respectively \cite{sikora2012,sikora2013,ftouni2015thermal}. Using these data of the SiN cantilever, we estimated the thermal conductance due to SiN material ($G=\kappa S/l$; $S$ cross-section area and $l$ is the length) to be approximately 10$^{-6}$~WK$^{-1}$. In addition, we took into account the four electrical gold leads with the following dimensions: length 200~$\mu$m, width 2~$\mu$m, and thickness 50~nm. The obtained thermal conductance due to the electrical leads is equivalent to $6\times 10^{-7}$~WK$^{-1}$. As all the thermal conductances are in parallel, they are additive, we then estimated the total thermal conductance $G_{probe}$ of the probe to be on the order of 1.6~$\times 10^{-6}$ WK$^{-1}$. 

This value is a little lower than the one measured above from the material composition of the probe $G_{probe}= 2.4\times 10^{-6}$~WK$^{-1}$; this can be explained by the fact that we omitted, in the calculation, the contribution to the thermal transport of the probe of the gold reflecting layer on the back side of the AFM probe.

\begin{figure}[!ht]
	\includegraphics[width=0.5\textwidth]{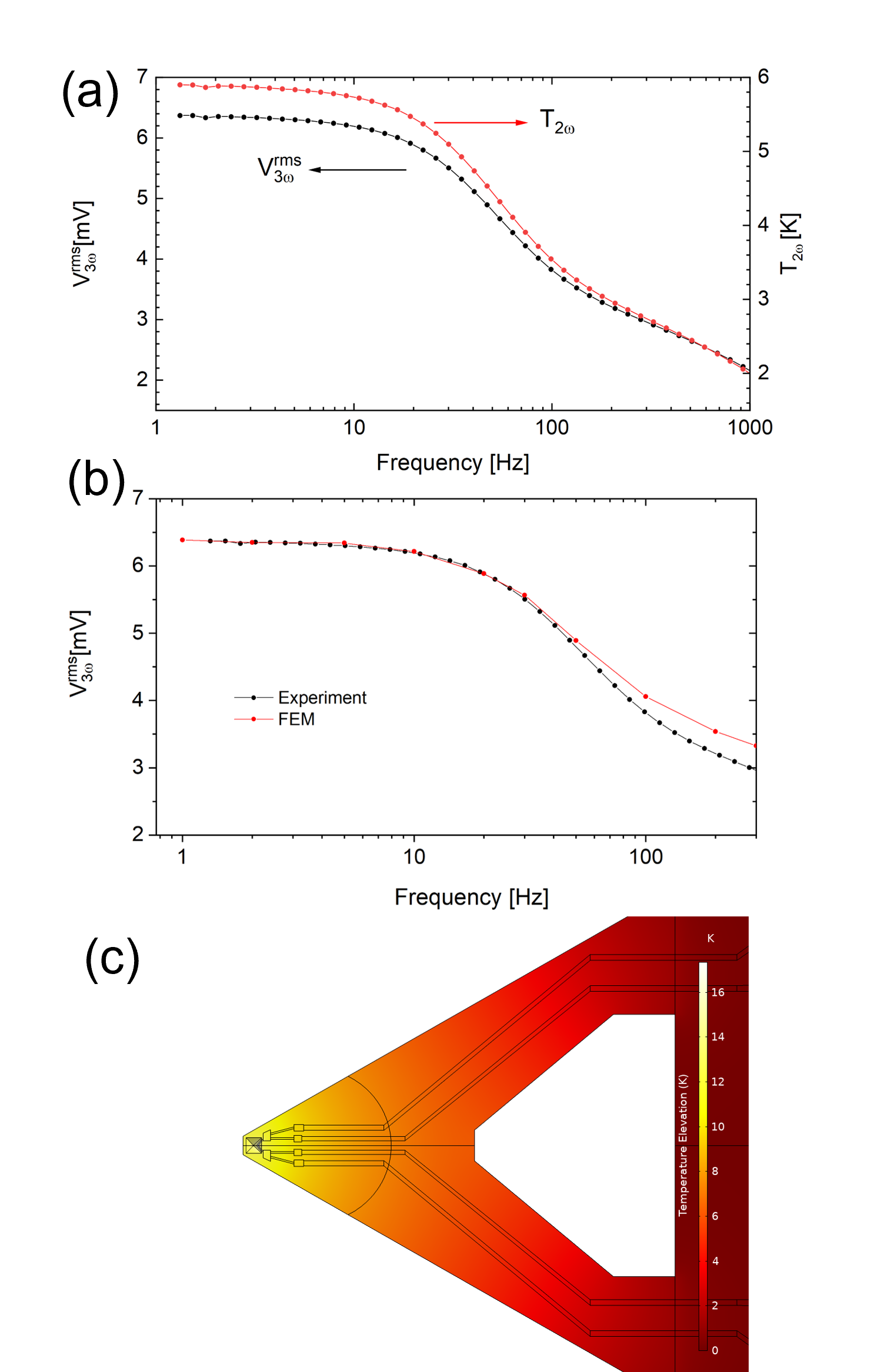}
	\caption{a) Measurement of the $V^{rms}_{3\omega}$ voltage amplitude and corresponding temperature oscillation amplitude of the SThM probe at 2$\omega$ as a function of electrical frequencies in vacuum conditions. b) The measured $V^{rms}_{3\omega}$ in black dots is plotted against the frequency whereas the red dots stands for the finite element method (FEM) simulation results. (c) Temperature profile in the probe cantilever as calculated by FEM, the temperature is given by the color scale that goes from 0~K (dark red) to 16~K (light yellow).}
	\label{Fig:8}
\end{figure}

In order to go beyond the approximate calculation of the probe’s thermal properties, we used the Finite Element Method (FEM) to numerically simulate the thermal behavior of the probe. A similar methodology as the one presented in ref.~\cite{pernot2021frequency} was carried out, where the 3D model constructed for this work follows the geometrical, electrical and thermal properties of the different materials of the NbN SThM probe as described in a previous work \cite{swami2022electron}. The model couples the electrical current with heat transfer module so that Joule heating is simulated in the probe by applying an electrical current as a boundary condition in the NbN layer. Concerning thermal boundary conditions, we assume that all free surfaces are adiabatic surfaces, this is consistent with the experiments done in vacuum where there is no convection loss. We can also neglect the heat flow through radiation since temperature gradients are small. The $3\omega$ voltage is obtained by post-processing the results of the temporal evolution of the voltage across the probe and isolating its $3\omega$ component from the first harmonic.

In Fig.~\ref{Fig:8}(b), for frequencies below the cut-off frequency, we report an excellent agreement between experimental results and simulations for both $3\omega$ rms voltage and frequency behavior. The discrepancy observed at frequencies above the cut-off frequency may be attributed to the slight deviations in the properties of the fabricated probe, especially its specific heat. The temperature profile along the probe is given in Fig.~\ref{Fig:8}(c) where the hot spot is clearly located at the apex of the pyramid.

The thermal time constant $\tau_{th}$ can be defined as the ratio of heat capacity to thermal conductance of the system: 
\begin{equation}
\tau_{th}=C_p/G_{probe}
\label{tau}
\end{equation}
where $C_p$ is the heat capacity with a unit of JK$^{-1}$. First, we wanted to calculate the mass of the heat source (as depicted in Figure~\ref{Fig:6}) located at the end of the cantilever; note that here we only considered the SiN material for calculation. The density of the SiN is taken from the literature which is 3~gcm$^{-1}$~\cite{ftouni2015thermal,tavakoli2022}. Second, the volume of the heat source is calculated to be 6.5~$\times 10^{-9}$ cm$^{3}$, using length, width and thickness of 30~$\mu$m, 20~$\mu$m and 0.8~$\mu$m, respectively corresponding to the isothermal triangular area of the cantilever (see Figure~\ref{Fig:6}). Finally, the mass of the heat source is found to be 1.5~ng. Then, the heat capacity $C_p$ can be calculated by multiplying the mass with the specific heat of the SiN and found to be $C_p=1.2~\times 10^{-9}$~JK$^{-1}$. Using Eq.~\ref{tau}, the thermal time constant is then estimated to be on the order of the millisecond $\tau_{th}=2~\times 10^{-3}$~s.

\subsection{Impact of the laser beam}

\begin{figure}[!ht]
\includegraphics[width=0.5\textwidth]{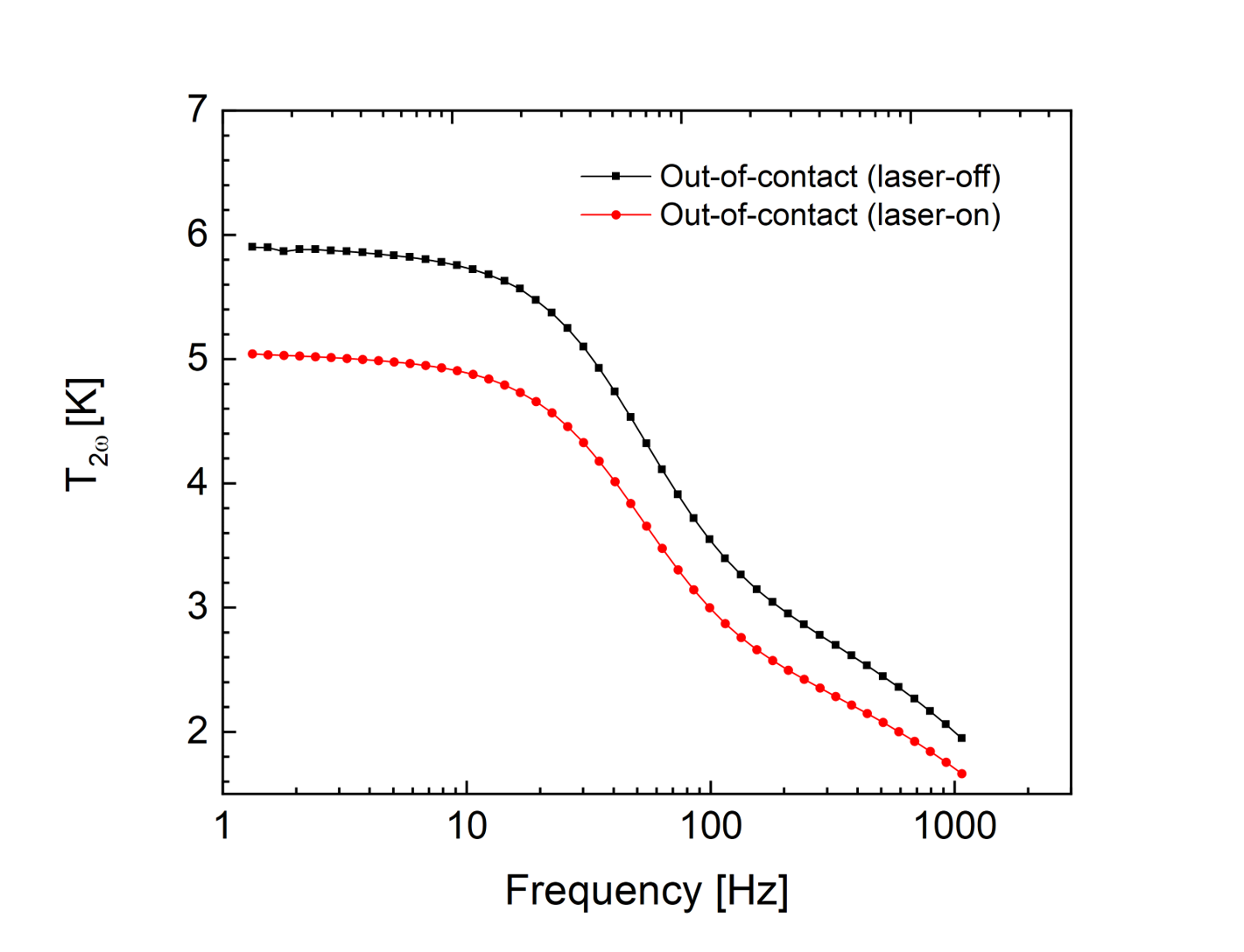}
	\caption{Temperature oscillation amplitude $ T_{2\omega}$ as a function of frequency in-presence and in-absence of the laser beam on the probe cantilever.}
	\label{Fig:9}
\end{figure}

When operating an AFM measurement, one needs to shine a laser beam on the cantilever to get the probe into contact with the sample. This is required to do thermal measurements with a controlled force between the probe and the sample or to perform force-distance curves. In this work, using the EnviroScop, the laser beam has a power of 1~mW knowing that only a fraction of the power will be absorbed by the cantilever. This power will contribute to raise the DC probe's temperature and modify its thermal response. In Figure~\ref{Fig:9}, the impact of the laser beam on the temperature oscillation amplitude is illustrated. This change can be explained by considering the DC overheating in the thermal probe via absorption of the laser beam. In our measurement, we have measured that this increase in the probe's DC temperature in turn decreases the electrical resistance $R_{0}$ and its TCR. $R_{0}$ of the NbN thermometer goes from 16~k$\Omega$ to 15~k$\Omega$ corresponding to a $\Delta T_{DC}=14$~K. Consequently, we observed the reduction in the voltage amplitude of the $V_{3\omega}$ in presence of the laser and the corresponding decrease in the amplitude of temperature oscillation $ T_{2\omega}$  as shown in Figures~\ref{Fig:10} (a) and (b). However, this change in the amplitude of $T_{2\omega}$ does not prevent further measurements in contact by $V_{3\omega}$-SThM experiment as we will see in the following.

In Figure~\ref{Fig:9}, when the laser illuminates the cantilever, the induced temperature oscillation amplitude in the probe is reduced from around 6~K to 5~K  for an electrical power dissipated in the SThM probe of $13.4\times10^{-6}$W by using an applied current of 30~$\mu$A in the electrical resistance of the NbN thermometer of 15~k$\Omega$. In this case, we measured a $G_{probe}$ to be $2.7\times10^{-6}$WK$^{-1}$, a higher value than the one measured from the laser-off experiment; it can be noted that the thermal time constant between on and off does not change. Since all the 3$\omega$-SThM measurements will be carried out in the presence of the laser, the precise knowledge of $G_{probe}$ in the laser-on configuration is essential to quantify the amount of heat exchanged between the probe and a sample surface.

\section{SThM measurements in contact and out-of-contact on sapphire sample}

After having demonstrated that we can obtain the thermal conductance of the cantilever from the $3\omega$ measurement, we will proceed with a thermal conductivity measurement using the SThM probe of a reference sample in contact, in this case a sapphire substrate, using the SThM probe operated in contact mode. Measurements were performed using an applied current of 30~$\mu$A ($\Delta T_{DC}<20$~K) to keep the sensor temperature close to ambient and therefore avoid any change in electrical properties due to temperature-dependent phenomena such as annealing of the thermometer or electrothermal migration.

In the previous section, we demonstrated that the thermal conductance of the cantilever of the probe is obtained from 3$\omega$ measurements. In this section, we proceed with the measurement of the thermal conductance of the probe-sample system, in this case the sample is crystalline sapphire substrate, operating the SThM in contact mode.
This measurement requires an appropriate thermal model. We will first discuss the details of the SThM thermal model developed to interpret the $V_{3 \omega}$ voltage, to understand the different routes of heat transport under vacuum conditions and to extract the thermal resistance or thermal conductance of the system. 

\subsection{Thermal model for thermal conductance measurement in contact}

\begin{figure}[!ht]
	\centering
	\includegraphics[width=0.5\textwidth]{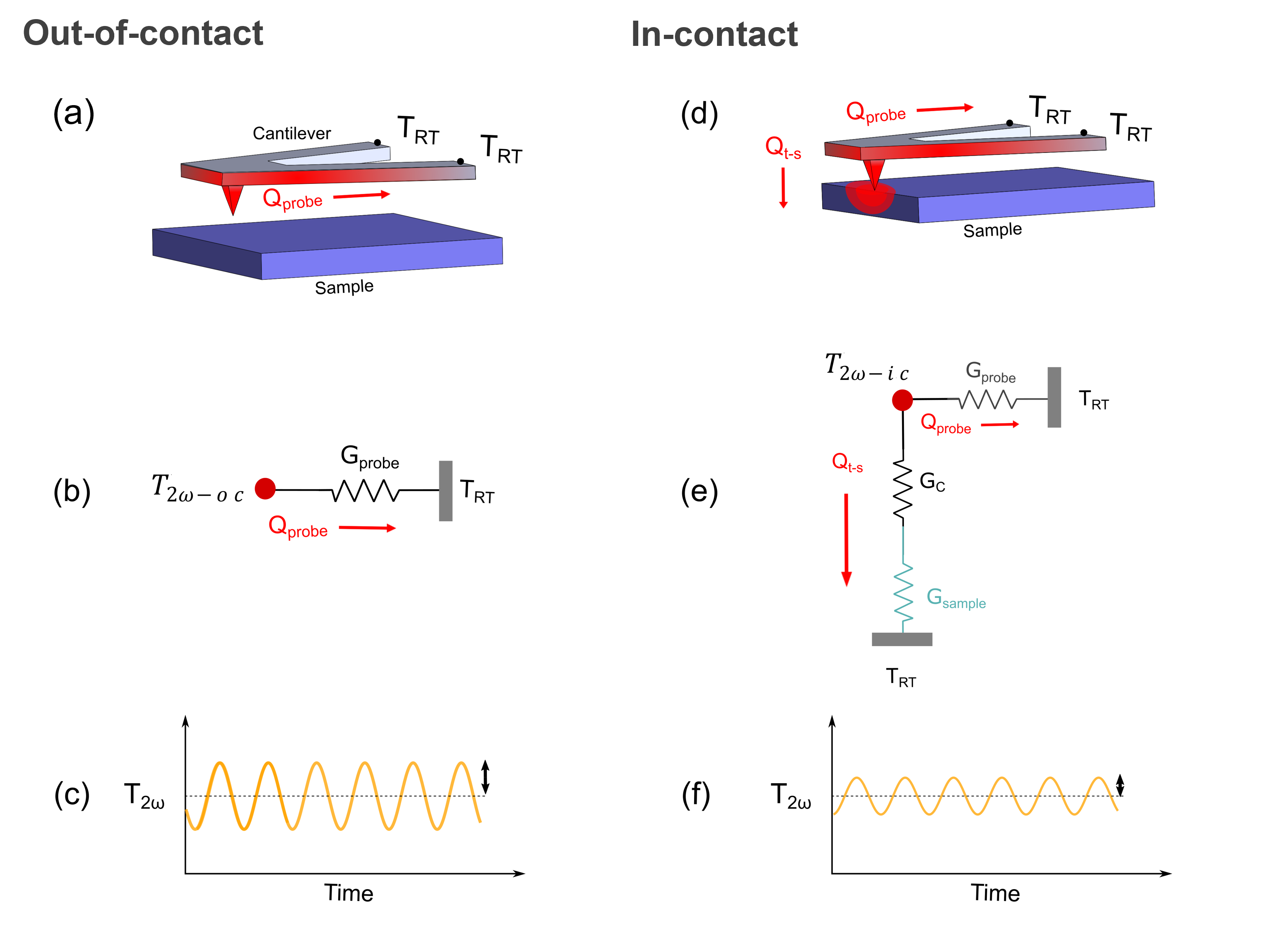}
	\caption{Comparison of the SThM thermal models between the out of contact (left) and the in-contact (right) regimes under vacuum conditions. (a) Schematic representation of the heat transfer through the probe and (b) the corresponding thermal model. (c) The sketch shows $ T_{2\omega-oc}$ temperature amplitude oscillation of the SThM probe (out-of-contact). (d) Schematic of heat transfer interactions between the probe and the sample. (e) the thermal model when the tip is in contact with the sample, and (f) the measured temperature amplitude $T_{2\omega-ic}$ response from the SThM probe.}
	\label{Fig:10}
\end{figure}

The comparison of the SThM thermal models between non-contact and in-contact is presented in Figure~\ref{Fig:10}, the probe being operated under vacuum in active 3$\omega$ mode in both cases. In the non-contact mode (see Figures~\ref{Fig:10} (a) to (c)), the heat transport can only occur from the tip to the chip body, which is represented by the heat flow $Q_{probe}$, the chip body acting as a heat sink at room temperature. In contact mode, Figures~\ref{Fig:10} (d) to (f)), results in opening another channel to heat flow from the probe into the sample. Due to the heat transport into the sample, the measured amplitude of temperature variation of the probe decreases, as shown in Figure~\ref{Fig:10} (f) as compared to Figure~\ref{Fig:10} (c). 

In case of contact, the additional thermal paths to $G_{probe}$ are of two kinds, firstly the thermal contact conductance between the tip and the surface $G_{c}$ and second the thermal conductance that corresponds to spreading the heat in the sample $G_{sample}$.

First, the thermal contact conductance $G_{c}$ will refer to the thermal conductance of the mechanical contact between the probe and the sample. $R_c =1/G_{c}$ naturally includes the thermal boundary resistance that occurs at the interface between two dissimilar materials or the same material with different crystallographic orientations that leads to temperature discontinuity at the interface. As a contact is never perfect, different parameters can influence $R_c$  \cite{gomes2015scanning,zhang2020review} such as (i) water, contamination or oxide layers that can cover surfaces and (ii) surface roughness or weak coupling bonds between the atoms of contacting solids. Advanced contact models have been developed to account for the weak coupling. The transmission probability is then related to the mechanical coupling spring between the two solids.   

$G_{sample}$ describes the conductance of heat flow into the material specimen under investigation. In case of a bulk homogeneous sample with an isotropic thermal conductivity $\kappa$ and whose dimensions are much larger than the dimensions of the heat source, the SThM probe’s hot apex can be modeled as a circular heat source at the flat apex of radius $b_c$. Then, the thermal spreading conductance of the sample can be expressed as \cite{yovanovich2003thermal}:
\begin{equation}
	G_{sample,spread} =  K \kappa b_c
	\label{thermalspreading}
\end{equation}
where $K$ is a geometrical factor describing the heat spreading within the sample, $\kappa$ the thermal conductivity of the sample and $b_{c}$ is the total contact radius that takes into account all the mechanisms of heat transfer through the solid-solid contact. $K=4$ is often used for a discoïdal contact as the heat source can be considered to be isothermal due to its small dimensions \cite{puyoo2010thermal}.
If $b_{c}$ is lower than the average mean free path of energy carriers in the material $\Lambda$, another thermal resistance within the sample must be considered at the level of the contact \cite{prasher2006}.  In order to account analytically for the partly ballistic dissipation in the substrate, we can rely on Wexler's approach \cite{wexler1966}, where thermal resistance is the sum of the diffusive and ballistic resistances \cite{nikolic1999}. It was shown numerically that the maximal error induced by this approach is 11 \% \cite{sharvin1965}. The ballistic thermal resistance, is given by \cite{wexler1966}:

\begin{equation}
	G_{sample,bal} = \frac{k}{\Lambda} \frac{3 \pi b_c^2}{4} 
		\label{equationbal}
\end{equation}

Since these conductances are in series then the total thermal conductance associated with the sample is given by:

\begin{equation}
\frac{1}{G_{sample}} = \frac{1}{G_{sample,spread}} + \frac{1}{G_{sample,bal}}
		\label{equgsample}
\end{equation}

Moreover, a thermal conductance between the NbN sensor and the tip apex ($G_{tip-apex}$) acts as an additional thermal conductance to $G_c$ and $G_{sample}$ since the NbN sensor is located on the pyramid and not at the tip apex as depicted in Fig.~\ref{Fig:2}~b.  The NbN thermometer being located  along one face of the square based pyramid (3~$\mu$m height, and 800~nm of SiN), the thermal conductance of NbN sensor and SiN pyramid $G_{tip-apex}$ is estimated to be on the order of $10^{-4}$~WK$^{-1}$, a thermal conductance much larger than any other thermal conductance of the probe (cantilever, contact, sample etc...); we will then neglect it in the following.

Note that $G_{tip-apex}$ could also include a contribution of ballistic heat conduction at the tip apex if the characteristic size at the tip apex and the contact ($b_c$) are smaller than the average mean free path of energy carriers in the tip’s material, specifically when the tip material is made of a crystalline semiconductor material. This is not the case for the amorphous SiN tip used in this work since the phonon mean free path for silicon nitride has been estimated to be smaller than the nanometer scale \cite{ftouni2015thermal}. 

The thermal conductance from the tip to the sample defined as $G_{t-s}$ is then the combination of two conductances in series, \textit{i.e.} the thermal contact conductance between the tip-sample $G_c$, and the thermal spreading conductance $G_{sample}$, $G_{t-s}$ can be written as:

\begin{equation}
	\frac{1}{G_{t-s}}= \frac{1}{G_{c}}+\frac{1}{G_{sample}}
\end{equation}

To establish the appropriate thermal modelling, we need to consider an experimental protocol with two different measurements of the temperature oscillation of the probe: 1-when the probe is out-of-contact with an oscillation temperature of $T_{2\omega-oc}$ and 2-when the probe is in-contact $T_{2\omega-ic}$; the indices $ic$ refer to in-contact mode.

Considering that the dissipated electrical power at the apex of the tip is identical to a first order in both cases when the probe is far from the sample or in contact, the balanced power equation between the total heat generated in the probe and the different routes of heat flux can be written as follows:

\begin{equation}
	P_{2\omega} = G_{probe} T_{2\omega-oc}
	\label{equationoc}
\end{equation}
when the tip is out-of-contact, and :
\begin{equation}
	P_{2\omega} = (G_{probe} + G_{t-s}) T_{2\omega-ic}
	\label{equationic}
\end{equation}
when the tip is in-contact with the sample. 

Then, from the Eq.~\ref{equationoc} and \ref{equationic} the thermal contact conductance $G_{c}$ or the sample conductance $G_{sample}$ can be directly expressed as a function of $G_{probe}$, known from out-of-contact measurement, and the two measured quantities $T_{2\omega-oc}$ and $T_{2\omega-ic}$, we then have either:

\begin{equation}
	G_{c}=\frac{G_{sample}}{\left ( \frac{G_{sample}}{G_{probe}} \left (\frac{ T_{2\omega-ic}}{ T_{2\omega-oc}- T_{2\omega-ic}}  \right ) - 1  \right ) } 
\label{equgc} 
\end{equation}
or
\begin{equation}
	G_{sample}=\frac{G_{c}}{\left ( \frac{G_{c}}{G_{probe}}  \left (\frac{ T_{2\omega-ic}}{ T_{2\omega-o.c}- T_{2\omega-ic}}  \right ) - 1  \right ) } 
\label{equgs} 
\end{equation}

In this work, where the tip is brought into contact with a bulk crystalline sapphire sample, we will utilize Eq.~\ref{equgc} to determine the thermal contact conductance $G_{c}$, a value of which the order of magnitude is essential for further study on unknown thermal conductivity materials.

\subsection{Thermal contact conductance at the solid-solid interface}

The mechanical contact between the SThM tip and the solid surface refers to the solid-solid interface. We already discussed that the heat flux through the heated tip into the solid sample depends on the thermal conductivity of the sample. This implies an increase in the thermal conductance associated to the heat amount transferred to the sample $G_{t-s}$ with increasing sample’s thermal conductivity, as proposed by Majumdar \cite{majumdar1999scanning}. However, the existence of a thermal contact conductance $G_{c}$ at the solid-solid interface limits the tip-sample thermal conductance $G_{t-s}$. The value of $G_{c}$ between the SThM tip and the sample surface will be the most important parameters that sets the sensitivity in thermal conductivity measurement of the SThM probe.

The following protocol is utilized to perform the $3\omega$-SThM measurement of the thermal contact conductance  $G_{c}$ of the probe to the sample. We carried out experiments on various locations on the sample surface with a constant applied force of few nanonewtons. Then the $V_{3\omega}$ signal is measured as a function of frequency in out-of-contact mode first that will serve as a reference and second in-contact mode. An average of the signal for each location is taken over twenty cycles of out-of-contact and in-contact experiments. Finally, the averaged $T_{2\omega}$ temperature is used to extract the value of the contact resistance using Eq.~\ref{equgc}, where all the other parameters are known such as $G_{sample}$ and $G_{probe}$.

\begin{figure}[!ht]
	\includegraphics[width=0.5\textwidth]{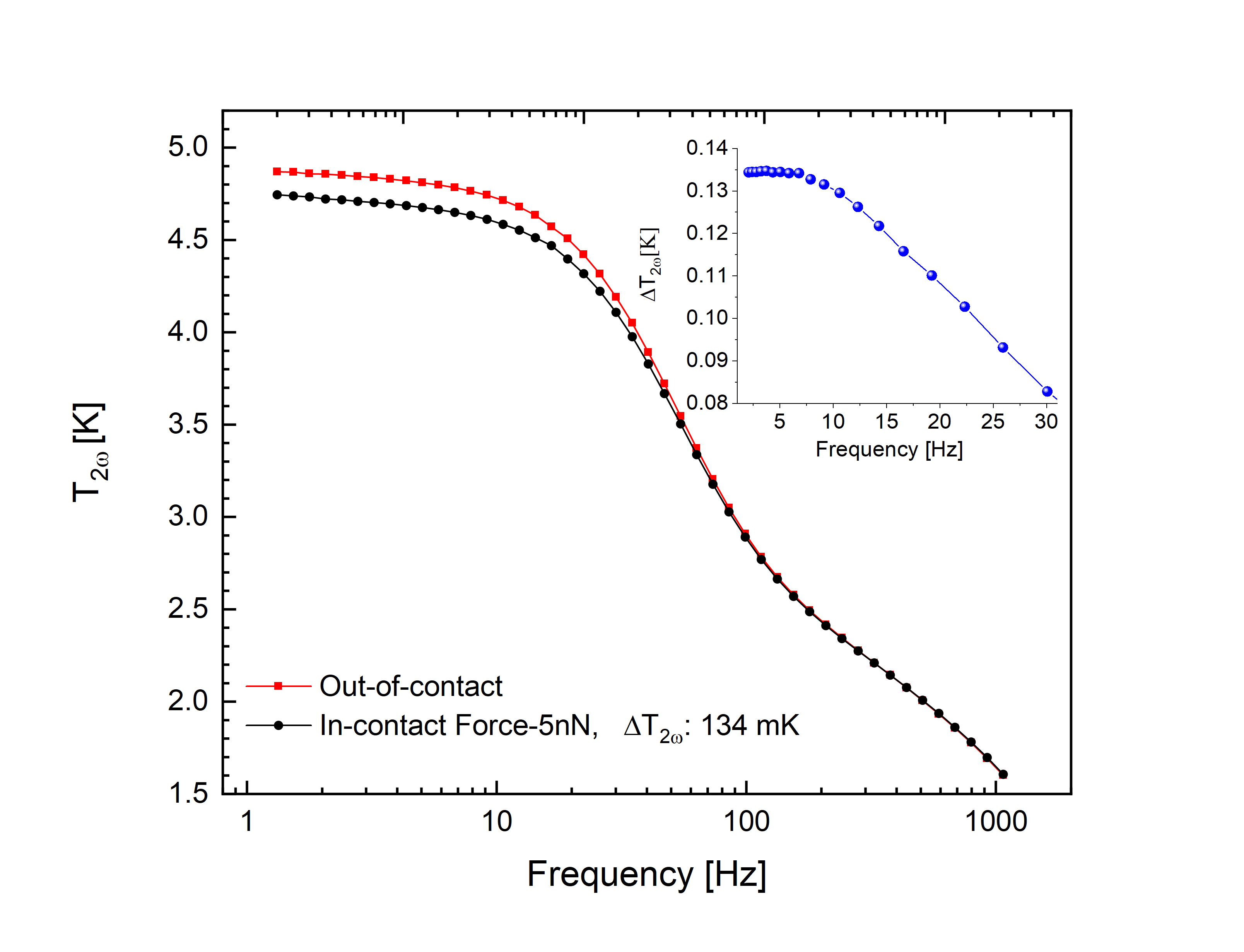}
	\caption{Averaged $T_{2\omega}$ amplitude oscillation of temperature measured versus frequency in out-of-contact and  in-contact configurations with a sapphire sample ($\kappa_{sapphire}$=34~Wm$^{-1}$K$^{-1}$) for an AC current of 30$\mu$A and an applied force of 5~nN. Red squares indicate the measured signal in the out-of-contact mode and red line is used to guide the eyes whereas black dots correspond to the in-contact measured temperature with blue line to guide the eyes. The inset shows the change of $T_{2\omega}$ between in-contact and out-of-contact mode, $\Delta T_{2\omega}$, at low frequency.}
	\label{Fig:11}
\end{figure}

We performed that SThM measurement of $G_{c}$ on a well-known crystalline sapphire samples. Figure~\ref{Fig:11} allows the comparison of the $ T_{2\omega}$ temperature oscillation of the SThM probe (extracted from $V_{3\omega}$ signal) in-contact and out-of-contact with the sapphire sample as a function of frequency. The red squares are the measured temperature oscillations of the probe in out-of-contact configuration corresponding to temperature $ T_{2\omega-oc}$ and black dots represent the measured temperature in-contact with the sample and correspond to $ T_{2\omega-ic}$. In the inset of Figure~\ref{Fig:11}, $\Delta T_{2\omega}=T_{2\omega-oc}-T_{2\omega-ic}$ is plotted showing a difference of 134~mK (at frequencies smaller than 10~Hz for an applied force of 5~nN) in the temperature oscillation between out-of-contact and in-contact configurations. 

\begin{figure}[!ht]
	\includegraphics[width=0.5\textwidth]{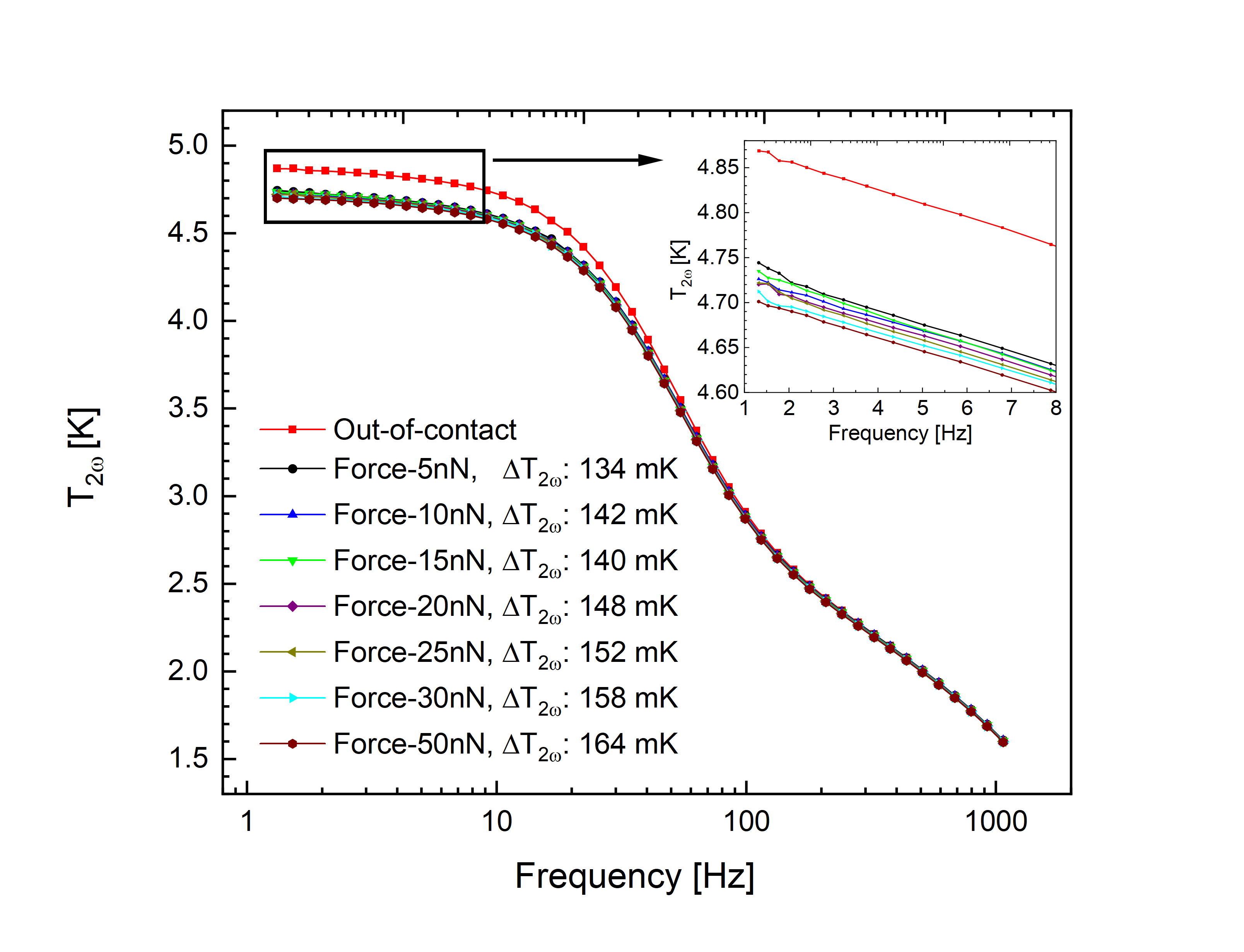}
	\caption{Averaged $T_{2\omega}$ amplitude oscillation of temperature measured versus frequency in out-of-contact and in-contact configurations with a sapphire sample ($\kappa_{sapphire}$=34~Wm$^{-1}$K$^{-1}$, AC current of 30$\mu$A) for different forces exerted on the cantilever. Red dots indicate the measured signal in the out-of-contact mode and color lines are used for the various applied forces from 5~nN to 50~nN.}
	\label{Fig:12}
\end{figure}

The same experiment has been done for different exerted forces, as it is shown Figure~\ref{Fig:12}. The inset of Figure~\ref{Fig:12} is clearly illustrating that this technique is sensitive enough to detect a change of $T_{2\omega}$ with an increased force. The thermal coupling is improved and consequently the change in the temperature oscillation $T_{2\omega}$ increases from 134~mK for 5~nN to 164~mK for 50~nN.
We can also notice that at high frequencies ($f>$40~Hz), the amplitude of the $T_{2\omega}$ temperature oscillations are similar for all configurations. 
That is because the probe behaves like a low-pass filter due to its thermal inertia. It means that the sample in contact is not probed anymore at high frequency, explaining the superposition of the $T_{2\omega}$ curves for frequencies greater than 40~Hz, a good indication of the quality of the measurement since, at these frequencies, only the probe is measured.

The thermal conductivity of the sapphire sample used in this work has been previously measured by regular 3$\omega$ method and found to be of 34~Wm$^{-1}$K$^{-1}$ on the very same material \cite{thesejessy,paterson2020}, knowing that the average mean free path of phonons in sapphire is about 130~nm \cite{hoogeboom2015}, it gives $G_{sample,spread}=1.7\times 10^{-5}$~WK$^{-1}$ and $G_{sample,bal}=9.6\times 10^{-6}$~WK$^{-1}$. From the equivalent radius of the tip apex measured by SEM to be 125~nm using the Eq.~\ref{equgc} we can estimate the $G_{sample}$ to be 6.15$\times 10^{-6}$~WK$^{-1}$. Using the experimental results obtained in this work for out-of-contact and in-contact $T_{2\omega}$ along with Eq.~\ref{equgc}, we can estimate the thermal conductance of the probe-sapphire contact to be $7.4\times 10^{-8}$~WK$^{-1}$, which can be translated into thermal boundary resistance per surface of $8.0\times 10^{-7}$~Km$^2$W$^{-1}$. Identical measurements were performed on silicon smooth surface of thermal conductivity of 148~Wm$^{-1}$K$^{-1}$ and a thermal conductance of the contact of $8.9 \times 10^{-8}$~WK$^{-1}$ has been measured, a value relatively closed to the one measured on sapphire.

These measurements of thermal boundary resistance can be compared to Pd probes (SiN probe) in vacuum ($P <10^{-3}$~mbar) having a curvature radius close to the NbN probe where a $G_{c}$ on the order of 1.4 to 16$\times 10^{-8}$~WK$^{-1}$ has been measured in the past, a value very close to $7.4\times 10^{-8}$~WK$^{-1}$ measured in this work \cite{pernot2021frequency,alikin2023,gonzalez2023,umatova2019}.

\subsection{Sensitivity function}
In order to characterize the performance of the NbN probe regarding the measurement of thermal conductivity of materials or thermal contact conductance, it is appropriate to calculate the dimensionless sensitivity functions defined as:
\begin{equation}
	S(G_{sample})=\frac{G_{sample}}{T_{2\omega-ic}}  \frac{d\Delta T_{2\omega}}{dG_{sample}} 
\label{equsens1} 
\end{equation}
Using Eq.~\ref{equgc}, we obtain :
\begin{equation}
\Delta T_{2\omega}=\frac{T_{2\omega-ic} G_{sample} G_{c}}{G_{p}(G_{sample}+ G_{c})}
\label{equsens2} 
\end{equation}
Replacing $G_{sample}$ by the simple spreading thermal conductance given by Eq.~\ref{thermalspreading} for a fixed thermal contact conductance $G_c$, we get for the sensitivity function expressed as a function of $\kappa$:
\begin{equation}
	S(\kappa)=\frac{4 b_c G_{c}^2 \kappa}{G_p (G_c+4b_c \kappa)^2}  
\label{equsens3} 
\end{equation}

Similarly, we can calculate the sensitivity function as a function of the contact conductance per surface $g_c=G_c/4b^2_c$, for a fixed thermal conductivity of sample $\kappa$:

\begin{equation}
	S(g_c)=\frac{4 b_c^2 \kappa^2 g_c}{G_p (\kappa+b_c g_c)^2}  
\label{equsens4} 
\end{equation}

\begin{figure}[!ht]
	\includegraphics[width=0.5\textwidth]{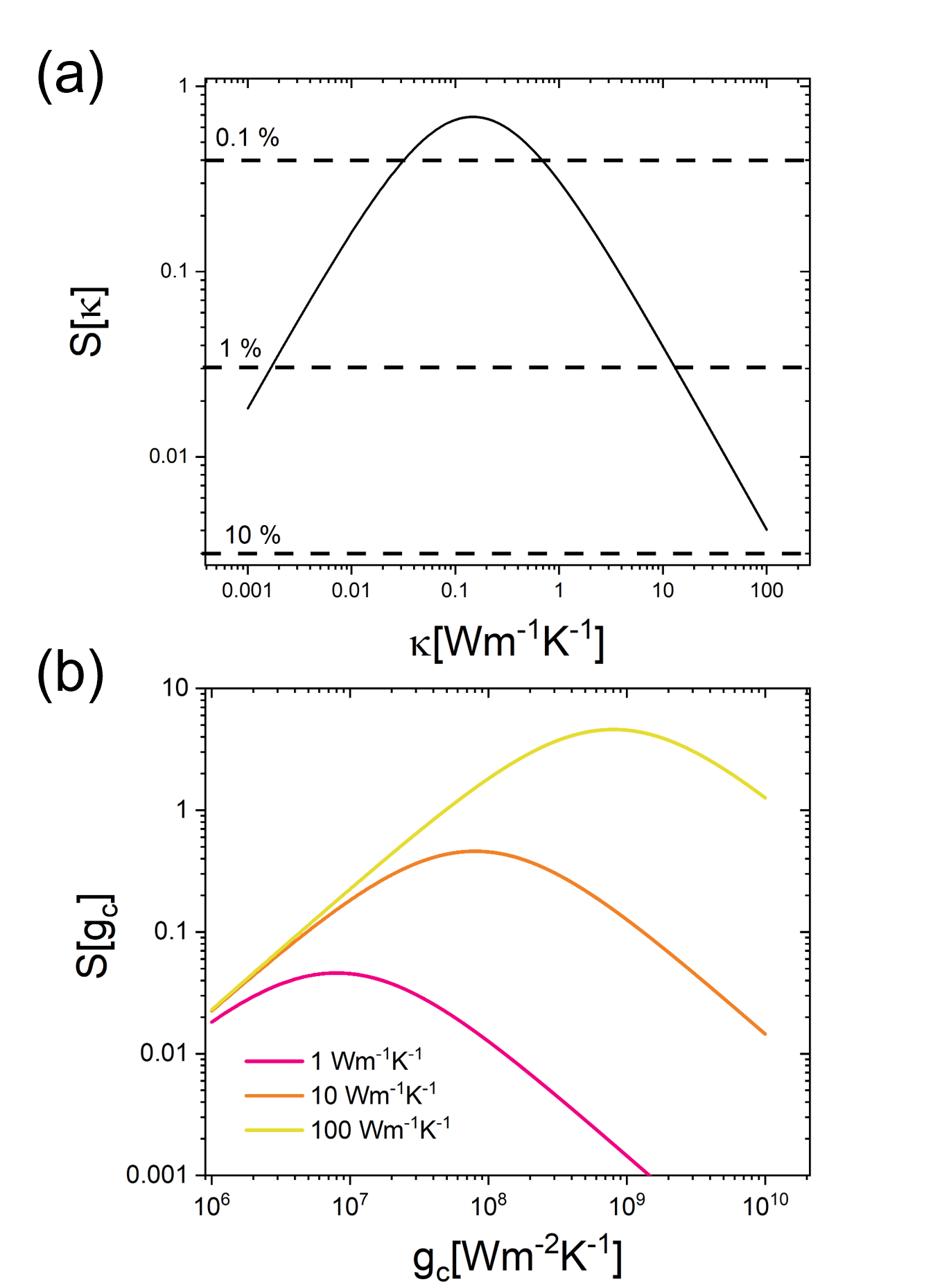}
	\caption{(a) Sensitivity function calculated from the probe parameters of this work as a function of the thermal conductivity of sample under study. The dashed lines illustrate the performances of the probe for different targeted percentages of sensitivity of the thermal conductivity $\Delta \kappa/\kappa$ calculated from the signal to noise ratio $V^{rms}_{3\omega}/noise$. (b) Sensitivity function as a function of the thermal contact conductance per surface for different thermal conductivity of samples : 1, 10 and 100~Wm$^{-1}$K$^{-1}$.}
	\label{Fig:13}
\end{figure}

In Fig.~\ref{Fig:13}, both sensitivity functions are shown. In Fig.~\ref{Fig:13}(a), $S(\kappa)$ is plotted as a function of the thermal conductivity $\kappa$ of the sample that will be probed during a SThM experiment (for a fixed thermal contact conductance $G_c$ as measured in this work on sapphire); the dashed lines indicate the performance of the probe for different targeted percentages of sensitivity of the thermal conductivity $\Delta \kappa/\kappa$ calculated from the signal to noise ratio $V^{rms}_{3\omega}/noise$ as obtained from Fig.~\ref{Fig:7}. In Fig.~\ref{Fig:13}(b) $S(g_c)$ is plotted as a function of the thermal contact conductance per surface for three different thermal conductivity of samples 1, 10 and 100~Wm$^{-1}$K$^{-1}$.

This shows that the probe characterized in this work has its optimum sensitivity for samples of thermal conductivity of 0.15~Wm$^{-1}$K$^{-1}$. However, if the experimental resolution targeted for the thermal conductance measurement defined as $\Delta \kappa/\kappa$ is above 1\% then a wide thermal conductivity range is accessible from 0.002 to 13~Wm$^{-1}$K$^{-1}$ as it is shown in Fig.~\ref{Fig:13}(a). Finally, if a lower tolerance is acceptable (above 10\%), measurements up to 100~Wm$^{-1}$K$^{-1}$ are possible, demonstrating the very wide thermal conductivity range of few orders of magnitude that is accessible with the NbN thermal probe having a blunt tip. Similarly, Fig.~\ref{Fig:13}(b) illustrates the potential performance of the NbN probe for studying thermal boundary resistance, with a particular sensitivity when measuring thermal contact on highly conductive samples (100~Wm$^{-1}$K$^{-1}$) like silicon for instance. This analysis is valid only for blunt probes. Further studies are required to estimate the sensitivity functions of probes with sharper tip.

\section{Conclusion}
In summary, we have presented the experimental set-up necessary to perform scanning thermal microscopy measurement using NbN thermal probes. The experimental environment including the AFM instrument, the probe holder, and the measurement chain has been described in details. The fabrication and the calibration of the SThM probe equipped with a highly sensitive NbN thermometer have been done in the AFM instrument. The present work demonstrates the potential of this experimental set-up coupled to highly sensitive NbN thermal probe for thermal transport measurement at the nanoscale. Thermal experiments based on the 3$\omega$ technique show the operability of these probes. We underline that we performed the experiments by inducing a temperature oscillation of only $\sim$5~K between the probe and the sample in 3$\omega$-SThM mode, which is significantly lower than the temperature variation used in most DC or AC heating active modes \cite{nelson2007measuring,zhang2012high,nguyen2019,bodzenta2020,pernot2021frequency}. 

The main advantages of this new SThM probe are based on the high temperature coefficient of the NbN sensor's electrical resistance, which offers a temperature resolution that has never been achieved in SThM. The use of small temperature differences can allow a greater stability of the probe over time by avoiding changes in its electrical properties due to temperature-dependent phenomena such as thermometer annealing for instance. Using small temperature differences will be particularly important when measuring samples having physical (mainly mechanical and thermal) properties that depend strongly on temperature such as phase change materials or biological systems. However, measurements in active DC mode with the NbN tip heated to several tens of Kelvin are still possible, which is particularly useful for studying the thermomechanical behavior of polymer materials for example. 

The NbN thermometers are more sensitive than those used in comparable thermal probes. However, the NbN thermometers are a little less sensitive on the AFM probe pyramid than the state-of-art NbN thermometers on regular flat surface obtained in our group showing a significant margin of progress for these probes in the near future \cite{bourgeois2006liquid}. 
Additionally, the response of the NbN thermometer on the AFM tip is measured using four-terminal wires. This allows better control of the response of the NbN sensor alone, since the electrical resistance of the connecting wires is not involved in the measurement. Moreover, the flat end of the tip is not coated with NbN, which limits mechanical wear on the thermal sensor and therefore ensures better measurement reproducibility.

The demonstration of such an instrument opens up numerous possibilities for high resolution thermal measurements at very small length scale, from the local temperature or thermal conductivity of composites and nanomaterials to the thermal conductance of nanometric-sized contacts. Finally, this technique based on NbN thermometry offers particularly exciting perspectives regarding SThM experiments in cryogenic conditions.

\textbf{Acknowledgement}
 The authors thank the technical support provided by Institut N\'eel, Nanofab for clean room fabrication of the probes, the Pole THEMA, especially P. Gasner and G. Moiroux for materials deposition, Electronic and the Pole Optique for AFM measurements. The authors acknowledge fruitful discussions with J.-L. Garden, B. Brisuda and F. Mazzelli. This work is supported by the Agence Nationale de la Recherche through the Tiptop project under grant agreement ANR-16-CE09-0023 and by SATT-Linksium through the project NANOTIP under grant agreement CM210012.

\textbf{Data Availability} Data are available upon request to the corresponding author.

\textbf{Supplementary Materials} The supplementary materials bring more data on the thermal conductance of the probe-sapphire contact, comparison with the thermal contact resistances found in literature and measured values in this work and finally the calculation from the measurement of the thermal resistance of the mechanical contact $R_c$ as a function of the force applied.

\textbf{Competing interests}
The authors declare no competing interests.

\bibliography{Swamisub2.bbl}

\begin{thebibliography}{68}%
\makeatletter
\providecommand \@ifxundefined [1]{%
 \@ifx{#1\undefined}
}%
\providecommand \@ifnum [1]{%
 \ifnum #1\expandafter \@firstoftwo
 \else \expandafter \@secondoftwo
 \fi
}%
\providecommand \@ifx [1]{%
 \ifx #1\expandafter \@firstoftwo
 \else \expandafter \@secondoftwo
 \fi
}%
\providecommand \natexlab [1]{#1}%
\providecommand \enquote  [1]{``#1''}%
\providecommand \bibnamefont  [1]{#1}%
\providecommand \bibfnamefont [1]{#1}%
\providecommand \citenamefont [1]{#1}%
\providecommand \href@noop [0]{\@secondoftwo}%
\providecommand \href [0]{\begingroup \@sanitize@url \@href}%
\providecommand \@href[1]{\@@startlink{#1}\@@href}%
\providecommand \@@href[1]{\endgroup#1\@@endlink}%
\providecommand \@sanitize@url [0]{\catcode `\\12\catcode `\$12\catcode
  `\&12\catcode `\#12\catcode `\^12\catcode `\_12\catcode `\%12\relax}%
\providecommand \@@startlink[1]{}%
\providecommand \@@endlink[0]{}%
\providecommand \url  [0]{\begingroup\@sanitize@url \@url }%
\providecommand \@url [1]{\endgroup\@href {#1}{\urlprefix }}%
\providecommand \urlprefix  [0]{URL }%
\providecommand \Eprint [0]{\href }%
\providecommand \doibase [0]{https://doi.org/}%
\providecommand \selectlanguage [0]{\@gobble}%
\providecommand \bibinfo  [0]{\@secondoftwo}%
\providecommand \bibfield  [0]{\@secondoftwo}%
\providecommand \translation [1]{[#1]}%
\providecommand \BibitemOpen [0]{}%
\providecommand \bibitemStop [0]{}%
\providecommand \bibitemNoStop [0]{.\EOS\space}%
\providecommand \EOS [0]{\spacefactor3000\relax}%
\providecommand \BibitemShut  [1]{\csname bibitem#1\endcsname}%
\let\auto@bib@innerbib\@empty
\bibitem [{\citenamefont {Pop}(2010)}]{pop2010energy}%
  \BibitemOpen
  \bibfield  {author} {\bibinfo {author} {\bibfnamefont {E.}~\bibnamefont
  {Pop}},\ }\bibfield  {title} {\bibinfo {title} {Energy dissipation and
  transport in nanoscale devices},\ }\href@noop {} {\bibfield  {journal}
  {\bibinfo  {journal} {Nano Res.}\ }\textbf {\bibinfo {volume} {3}},\ \bibinfo
  {pages} {147} (\bibinfo {year} {2010})}\BibitemShut {NoStop}%
\bibitem [{\citenamefont {Bourgeois}\ \emph {et~al.}(2016)\citenamefont
  {Bourgeois}, \citenamefont {Tainoff}, \citenamefont {Tavakoli}, \citenamefont
  {Liu}, \citenamefont {Blanc}, \citenamefont {Boukhari}, \citenamefont
  {Barski},\ and\ \citenamefont {Hadji}}]{bourgeois2016reduction}%
  \BibitemOpen
  \bibfield  {author} {\bibinfo {author} {\bibfnamefont {O.}~\bibnamefont
  {Bourgeois}}, \bibinfo {author} {\bibfnamefont {D.}~\bibnamefont {Tainoff}},
  \bibinfo {author} {\bibfnamefont {A.}~\bibnamefont {Tavakoli}}, \bibinfo
  {author} {\bibfnamefont {Y.}~\bibnamefont {Liu}}, \bibinfo {author}
  {\bibfnamefont {C.}~\bibnamefont {Blanc}}, \bibinfo {author} {\bibfnamefont
  {M.}~\bibnamefont {Boukhari}}, \bibinfo {author} {\bibfnamefont
  {A.}~\bibnamefont {Barski}},\ and\ \bibinfo {author} {\bibfnamefont
  {E.}~\bibnamefont {Hadji}},\ }\bibfield  {title} {\bibinfo {title} {Reduction
  of phonon mean free path: From low-temperature physics to room temperature
  applications in thermoelectricity},\ }\href@noop {} {\bibfield  {journal}
  {\bibinfo  {journal} {C. R. Phys.}\ }\textbf {\bibinfo {volume} {17}},\
  \bibinfo {pages} {1154} (\bibinfo {year} {2016})}\BibitemShut {NoStop}%
\bibitem [{\citenamefont {Zardo}\ and\ \citenamefont
  {Rurali}(2019)}]{zardo2019manipulating}%
  \BibitemOpen
  \bibfield  {author} {\bibinfo {author} {\bibfnamefont {I.}~\bibnamefont
  {Zardo}}\ and\ \bibinfo {author} {\bibfnamefont {R.}~\bibnamefont {Rurali}},\
  }\bibfield  {title} {\bibinfo {title} {Manipulating phonons at the nanoscale:
  Impurities and boundaries},\ }\href@noop {} {\bibfield  {journal} {\bibinfo
  {journal} {Curr. Opin. Green Sustain. Chem.}\ }\textbf {\bibinfo {volume}
  {17}},\ \bibinfo {pages} {1} (\bibinfo {year} {2019})}\BibitemShut {NoStop}%
\bibitem [{\citenamefont {Moore}\ and\ \citenamefont
  {Shi}(2014)}]{moore2014emerging}%
  \BibitemOpen
  \bibfield  {author} {\bibinfo {author} {\bibfnamefont {A.~L.}\ \bibnamefont
  {Moore}}\ and\ \bibinfo {author} {\bibfnamefont {L.}~\bibnamefont {Shi}},\
  }\bibfield  {title} {\bibinfo {title} {Emerging challenges and materials for
  thermal management of electronics},\ }\href@noop {} {\bibfield  {journal}
  {\bibinfo  {journal} {Mater. today}\ }\textbf {\bibinfo {volume} {17}},\
  \bibinfo {pages} {163} (\bibinfo {year} {2014})}\BibitemShut {NoStop}%
\bibitem [{\citenamefont {D{\'a}vila}\ \emph {et~al.}(2012)\citenamefont
  {D{\'a}vila}, \citenamefont {Tarancon}, \citenamefont {Calaza}, \citenamefont
  {Salleras}, \citenamefont {Fern{\'a}ndez-Reg{\'u}lez}, \citenamefont
  {San~Paulo},\ and\ \citenamefont {Fonseca}}]{davila2012monolithically}%
  \BibitemOpen
  \bibfield  {author} {\bibinfo {author} {\bibfnamefont {D.}~\bibnamefont
  {D{\'a}vila}}, \bibinfo {author} {\bibfnamefont {A.}~\bibnamefont
  {Tarancon}}, \bibinfo {author} {\bibfnamefont {C.}~\bibnamefont {Calaza}},
  \bibinfo {author} {\bibfnamefont {M.}~\bibnamefont {Salleras}}, \bibinfo
  {author} {\bibfnamefont {M.}~\bibnamefont {Fern{\'a}ndez-Reg{\'u}lez}},
  \bibinfo {author} {\bibfnamefont {A.}~\bibnamefont {San~Paulo}},\ and\
  \bibinfo {author} {\bibfnamefont {L.}~\bibnamefont {Fonseca}},\ }\bibfield
  {title} {\bibinfo {title} {Monolithically integrated thermoelectric energy
  harvester based on silicon nanowire arrays for powering micro/nanodevices},\
  }\href@noop {} {\bibfield  {journal} {\bibinfo  {journal} {Nano Energy}\
  }\textbf {\bibinfo {volume} {1}},\ \bibinfo {pages} {812} (\bibinfo {year}
  {2012})}\BibitemShut {NoStop}%
\bibitem [{\citenamefont {Perez-Mar{\'\i}n}\ \emph {et~al.}(2014)\citenamefont
  {Perez-Mar{\'\i}n}, \citenamefont {Lopeand{\'\i}a}, \citenamefont {Abad},
  \citenamefont {Ferrando-Villaba}, \citenamefont {Garcia}, \citenamefont
  {Lopez}, \citenamefont {Mu{\~n}oz-Pascual},\ and\ \citenamefont
  {Rodr{\'\i}guez-Viejo}}]{perez2014micropower}%
  \BibitemOpen
  \bibfield  {author} {\bibinfo {author} {\bibfnamefont {A.}~\bibnamefont
  {Perez-Mar{\'\i}n}}, \bibinfo {author} {\bibfnamefont {A.}~\bibnamefont
  {Lopeand{\'\i}a}}, \bibinfo {author} {\bibfnamefont {L.}~\bibnamefont
  {Abad}}, \bibinfo {author} {\bibfnamefont {P.}~\bibnamefont
  {Ferrando-Villaba}}, \bibinfo {author} {\bibfnamefont {G.}~\bibnamefont
  {Garcia}}, \bibinfo {author} {\bibfnamefont {A.}~\bibnamefont {Lopez}},
  \bibinfo {author} {\bibfnamefont {F.}~\bibnamefont {Mu{\~n}oz-Pascual}},\
  and\ \bibinfo {author} {\bibfnamefont {J.}~\bibnamefont
  {Rodr{\'\i}guez-Viejo}},\ }\bibfield  {title} {\bibinfo {title} {Micropower
  thermoelectric generator from thin si membranes},\ }\href@noop {} {\bibfield
  {journal} {\bibinfo  {journal} {Nano Energy}\ }\textbf {\bibinfo {volume}
  {4}},\ \bibinfo {pages} {73} (\bibinfo {year} {2014})}\BibitemShut {NoStop}%
\bibitem [{\citenamefont {Varpula}\ \emph {et~al.}(2017)\citenamefont
  {Varpula}, \citenamefont {Timofeev}, \citenamefont {Shchepetov},
  \citenamefont {Grigoras}, \citenamefont {Hassel}, \citenamefont {Ahopelto},
  \citenamefont {Ylilammi},\ and\ \citenamefont
  {Prunnila}}]{varpula2017thermoelectric}%
  \BibitemOpen
  \bibfield  {author} {\bibinfo {author} {\bibfnamefont {A.}~\bibnamefont
  {Varpula}}, \bibinfo {author} {\bibfnamefont {A.~V.}\ \bibnamefont
  {Timofeev}}, \bibinfo {author} {\bibfnamefont {A.}~\bibnamefont
  {Shchepetov}}, \bibinfo {author} {\bibfnamefont {K.}~\bibnamefont
  {Grigoras}}, \bibinfo {author} {\bibfnamefont {J.}~\bibnamefont {Hassel}},
  \bibinfo {author} {\bibfnamefont {J.}~\bibnamefont {Ahopelto}}, \bibinfo
  {author} {\bibfnamefont {M.}~\bibnamefont {Ylilammi}},\ and\ \bibinfo
  {author} {\bibfnamefont {M.}~\bibnamefont {Prunnila}},\ }\bibfield  {title}
  {\bibinfo {title} {Thermoelectric thermal detectors based on ultra-thin
  heavily doped single-crystal silicon membranes},\ }\href@noop {} {\bibfield
  {journal} {\bibinfo  {journal} {Appl. Phys. Lett.}\ }\textbf {\bibinfo
  {volume} {110}},\ \bibinfo {pages} {262101} (\bibinfo {year}
  {2017})}\BibitemShut {NoStop}%
\bibitem [{\citenamefont {Tainoff}\ \emph {et~al.}(2019)\citenamefont
  {Tainoff}, \citenamefont {Proudhom}, \citenamefont {Tur}, \citenamefont
  {Crozes}, \citenamefont {Dufresnes}, \citenamefont {Dumont}, \citenamefont
  {Bourgault},\ and\ \citenamefont {Bourgeois}}]{tainoff2019network}%
  \BibitemOpen
  \bibfield  {author} {\bibinfo {author} {\bibfnamefont {D.}~\bibnamefont
  {Tainoff}}, \bibinfo {author} {\bibfnamefont {A.}~\bibnamefont {Proudhom}},
  \bibinfo {author} {\bibfnamefont {C.}~\bibnamefont {Tur}}, \bibinfo {author}
  {\bibfnamefont {T.}~\bibnamefont {Crozes}}, \bibinfo {author} {\bibfnamefont
  {S.}~\bibnamefont {Dufresnes}}, \bibinfo {author} {\bibfnamefont
  {S.}~\bibnamefont {Dumont}}, \bibinfo {author} {\bibfnamefont
  {D.}~\bibnamefont {Bourgault}},\ and\ \bibinfo {author} {\bibfnamefont
  {O.}~\bibnamefont {Bourgeois}},\ }\bibfield  {title} {\bibinfo {title}
  {Network of thermoelectric nanogenerators for low power energy harvesting},\
  }\href@noop {} {\bibfield  {journal} {\bibinfo  {journal} {Nano Energy}\
  }\textbf {\bibinfo {volume} {57}},\ \bibinfo {pages} {804} (\bibinfo {year}
  {2019})}\BibitemShut {NoStop}%
\bibitem [{\citenamefont {Singhal}\ \emph {et~al.}(2019)\citenamefont
  {Singhal}, \citenamefont {Paterson}, \citenamefont {Ben-Khedim},
  \citenamefont {Tainoff}, \citenamefont {Cagnon}, \citenamefont {Richard},
  \citenamefont {Chavez-Angel}, \citenamefont {Fernandez}, \citenamefont
  {Sotomayor-Torres}, \citenamefont {Lacroix}, \citenamefont {Bourgault},
  \citenamefont {Buttard},\ and\ \citenamefont
  {Bourgeois}}]{singhal2019nanowire}%
  \BibitemOpen
  \bibfield  {author} {\bibinfo {author} {\bibfnamefont {D.}~\bibnamefont
  {Singhal}}, \bibinfo {author} {\bibfnamefont {J.}~\bibnamefont {Paterson}},
  \bibinfo {author} {\bibfnamefont {M.}~\bibnamefont {Ben-Khedim}}, \bibinfo
  {author} {\bibfnamefont {D.}~\bibnamefont {Tainoff}}, \bibinfo {author}
  {\bibfnamefont {L.}~\bibnamefont {Cagnon}}, \bibinfo {author} {\bibfnamefont
  {J.}~\bibnamefont {Richard}}, \bibinfo {author} {\bibfnamefont
  {E.}~\bibnamefont {Chavez-Angel}}, \bibinfo {author} {\bibfnamefont {J.~J.}\
  \bibnamefont {Fernandez}}, \bibinfo {author} {\bibfnamefont {C.~M.}\
  \bibnamefont {Sotomayor-Torres}}, \bibinfo {author} {\bibfnamefont
  {D.}~\bibnamefont {Lacroix}}, \bibinfo {author} {\bibfnamefont
  {D.}~\bibnamefont {Bourgault}}, \bibinfo {author} {\bibfnamefont
  {D.}~\bibnamefont {Buttard}},\ and\ \bibinfo {author} {\bibfnamefont
  {O.}~\bibnamefont {Bourgeois}},\ }\bibfield  {title} {\bibinfo {title}
  {Nanowire forest of pnictogen--chalcogenide alloys for thermoelectricity},\
  }\href@noop {} {\bibfield  {journal} {\bibinfo  {journal} {Nanoscale}\
  }\textbf {\bibinfo {volume} {11}},\ \bibinfo {pages} {13423} (\bibinfo {year}
  {2019})}\BibitemShut {NoStop}%
\bibitem [{\citenamefont {Harzheim}\ \emph {et~al.}(2020)\citenamefont
  {Harzheim}, \citenamefont {Evangeli}, \citenamefont {Kolosov},\ and\
  \citenamefont {Gehring}}]{harzheim2020direct}%
  \BibitemOpen
  \bibfield  {author} {\bibinfo {author} {\bibfnamefont {A.}~\bibnamefont
  {Harzheim}}, \bibinfo {author} {\bibfnamefont {C.}~\bibnamefont {Evangeli}},
  \bibinfo {author} {\bibfnamefont {O.~V.}\ \bibnamefont {Kolosov}},\ and\
  \bibinfo {author} {\bibfnamefont {P.}~\bibnamefont {Gehring}},\ }\bibfield
  {title} {\bibinfo {title} {Direct mapping of local seebeck coefficient in 2d
  material nanostructures via scanning thermal gate microscopy},\ }\href@noop
  {} {\bibfield  {journal} {\bibinfo  {journal} {2D Materials}\ }\textbf
  {\bibinfo {volume} {7}},\ \bibinfo {pages} {041004} (\bibinfo {year}
  {2020})}\BibitemShut {NoStop}%
\bibitem [{\citenamefont {Menges}\ \emph {et~al.}(2016)\citenamefont {Menges},
  \citenamefont {Riel}, \citenamefont {Stemmer},\ and\ \citenamefont
  {Gotsmann}}]{menges2016nanoscale}%
  \BibitemOpen
  \bibfield  {author} {\bibinfo {author} {\bibfnamefont {F.}~\bibnamefont
  {Menges}}, \bibinfo {author} {\bibfnamefont {H.}~\bibnamefont {Riel}},
  \bibinfo {author} {\bibfnamefont {A.}~\bibnamefont {Stemmer}},\ and\ \bibinfo
  {author} {\bibfnamefont {B.}~\bibnamefont {Gotsmann}},\ }\bibfield  {title}
  {\bibinfo {title} {Nanoscale thermometry by scanning thermal microscopy},\
  }\href@noop {} {\bibfield  {journal} {\bibinfo  {journal} {Rev. Sci.
  Instrum.}\ }\textbf {\bibinfo {volume} {87}},\ \bibinfo {pages} {074902}
  (\bibinfo {year} {2016})}\BibitemShut {NoStop}%
\bibitem [{\citenamefont {Tavakoli}\ \emph {et~al.}(2018)\citenamefont
  {Tavakoli}, \citenamefont {Lulla}, \citenamefont {Crozes}, \citenamefont
  {Mingo}, \citenamefont {Collin},\ and\ \citenamefont
  {Bourgeois}}]{tavakoli2018}%
  \BibitemOpen
  \bibfield  {author} {\bibinfo {author} {\bibfnamefont {A.}~\bibnamefont
  {Tavakoli}}, \bibinfo {author} {\bibfnamefont {K.}~\bibnamefont {Lulla}},
  \bibinfo {author} {\bibfnamefont {T.}~\bibnamefont {Crozes}}, \bibinfo
  {author} {\bibfnamefont {N.}~\bibnamefont {Mingo}}, \bibinfo {author}
  {\bibfnamefont {E.}~\bibnamefont {Collin}},\ and\ \bibinfo {author}
  {\bibfnamefont {O.}~\bibnamefont {Bourgeois}},\ }\bibfield  {title} {\bibinfo
  {title} {Heat conduction measurements in ballistic 1d phonon waveguides
  indicate breakdown of the thermal conductance quantization},\ }\href@noop {}
  {\bibfield  {journal} {\bibinfo  {journal} {Nat. Commun.}\ }\textbf {\bibinfo
  {volume} {9}},\ \bibinfo {pages} {4287} (\bibinfo {year} {2018})}\BibitemShut
  {NoStop}%
\bibitem [{\citenamefont {Cui}\ \emph {et~al.}(2019)\citenamefont {Cui},
  \citenamefont {Hur}, \citenamefont {Akbar}, \citenamefont {Kl{\"o}ckner},
  \citenamefont {Jeong}, \citenamefont {Pauly}, \citenamefont {Jang},
  \citenamefont {Reddy},\ and\ \citenamefont {Meyhofer}}]{cui2019thermal}%
  \BibitemOpen
  \bibfield  {author} {\bibinfo {author} {\bibfnamefont {L.}~\bibnamefont
  {Cui}}, \bibinfo {author} {\bibfnamefont {S.}~\bibnamefont {Hur}}, \bibinfo
  {author} {\bibfnamefont {Z.~A.}\ \bibnamefont {Akbar}}, \bibinfo {author}
  {\bibfnamefont {J.~C.}\ \bibnamefont {Kl{\"o}ckner}}, \bibinfo {author}
  {\bibfnamefont {W.}~\bibnamefont {Jeong}}, \bibinfo {author} {\bibfnamefont
  {F.}~\bibnamefont {Pauly}}, \bibinfo {author} {\bibfnamefont {S.-Y.}\
  \bibnamefont {Jang}}, \bibinfo {author} {\bibfnamefont {P.}~\bibnamefont
  {Reddy}},\ and\ \bibinfo {author} {\bibfnamefont {E.}~\bibnamefont
  {Meyhofer}},\ }\bibfield  {title} {\bibinfo {title} {Thermal conductance of
  single-molecule junctions},\ }\href@noop {} {\bibfield  {journal} {\bibinfo
  {journal} {Nature}\ }\textbf {\bibinfo {volume} {572}},\ \bibinfo {pages}
  {628} (\bibinfo {year} {2019})}\BibitemShut {NoStop}%
\bibitem [{\citenamefont {Gehring}\ \emph {et~al.}(2019)\citenamefont
  {Gehring}, \citenamefont {Van Der~Star}, \citenamefont {Evangeli},
  \citenamefont {Le~Roy}, \citenamefont {Bogani}, \citenamefont {Kolosov},\
  and\ \citenamefont {Van Der~Zant}}]{gehring2019efficient}%
  \BibitemOpen
  \bibfield  {author} {\bibinfo {author} {\bibfnamefont {P.}~\bibnamefont
  {Gehring}}, \bibinfo {author} {\bibfnamefont {M.}~\bibnamefont {Van
  Der~Star}}, \bibinfo {author} {\bibfnamefont {C.}~\bibnamefont {Evangeli}},
  \bibinfo {author} {\bibfnamefont {J.~J.}\ \bibnamefont {Le~Roy}}, \bibinfo
  {author} {\bibfnamefont {L.}~\bibnamefont {Bogani}}, \bibinfo {author}
  {\bibfnamefont {O.~V.}\ \bibnamefont {Kolosov}},\ and\ \bibinfo {author}
  {\bibfnamefont {H.~S.}\ \bibnamefont {Van Der~Zant}},\ }\bibfield  {title}
  {\bibinfo {title} {Efficient heating of single-molecule junctions for
  thermoelectric studies at cryogenic temperatures},\ }\href@noop {} {\bibfield
   {journal} {\bibinfo  {journal} {Appl. Phys. Lett.}\ }\textbf {\bibinfo
  {volume} {115}},\ \bibinfo {pages} {073103} (\bibinfo {year}
  {2019})}\BibitemShut {NoStop}%
\bibitem [{\citenamefont {Cahill}\ \emph {et~al.}(2014)\citenamefont {Cahill},
  \citenamefont {Braun}, \citenamefont {Chen}, \citenamefont {Clarke},
  \citenamefont {Fan}, \citenamefont {Goodson}, \citenamefont {Keblinski},
  \citenamefont {King}, \citenamefont {Mahan}, \citenamefont {Majumdar},
  \citenamefont {Maris}, \citenamefont {Phillpot}, \citenamefont {Pop},\ and\
  \citenamefont {Shi}}]{cahill2014}%
  \BibitemOpen
  \bibfield  {author} {\bibinfo {author} {\bibfnamefont {D.~G.}\ \bibnamefont
  {Cahill}}, \bibinfo {author} {\bibfnamefont {P.~V.}\ \bibnamefont {Braun}},
  \bibinfo {author} {\bibfnamefont {G.}~\bibnamefont {Chen}}, \bibinfo {author}
  {\bibfnamefont {D.~R.}\ \bibnamefont {Clarke}}, \bibinfo {author}
  {\bibfnamefont {S.}~\bibnamefont {Fan}}, \bibinfo {author} {\bibfnamefont
  {K.~E.}\ \bibnamefont {Goodson}}, \bibinfo {author} {\bibfnamefont
  {P.}~\bibnamefont {Keblinski}}, \bibinfo {author} {\bibfnamefont {W.~P.}\
  \bibnamefont {King}}, \bibinfo {author} {\bibfnamefont {G.~D.}\ \bibnamefont
  {Mahan}}, \bibinfo {author} {\bibfnamefont {A.}~\bibnamefont {Majumdar}},
  \bibinfo {author} {\bibfnamefont {H.~J.}\ \bibnamefont {Maris}}, \bibinfo
  {author} {\bibfnamefont {S.~R.}\ \bibnamefont {Phillpot}}, \bibinfo {author}
  {\bibfnamefont {E.}~\bibnamefont {Pop}},\ and\ \bibinfo {author}
  {\bibfnamefont {L.}~\bibnamefont {Shi}},\ }\bibfield  {title} {\bibinfo
  {title} {Nanoscale thermal transport. ii. 2003-2012},\ }\href
  {https://doi.org/10.1063/1.4832615} {\bibfield  {journal} {\bibinfo
  {journal} {Appl. Phys. Rev.}\ }\textbf {\bibinfo {volume} {1}},\ \bibinfo
  {pages} {011305} (\bibinfo {year} {2014})}\BibitemShut {NoStop}%
\bibitem [{\citenamefont {Zhao}\ \emph {et~al.}(2016)\citenamefont {Zhao},
  \citenamefont {Qian}, \citenamefont {Gu}, \citenamefont {Jajja},\ and\
  \citenamefont {Yang}}]{zhao2016measurement}%
  \BibitemOpen
  \bibfield  {author} {\bibinfo {author} {\bibfnamefont {D.}~\bibnamefont
  {Zhao}}, \bibinfo {author} {\bibfnamefont {X.}~\bibnamefont {Qian}}, \bibinfo
  {author} {\bibfnamefont {X.}~\bibnamefont {Gu}}, \bibinfo {author}
  {\bibfnamefont {S.~A.}\ \bibnamefont {Jajja}},\ and\ \bibinfo {author}
  {\bibfnamefont {R.}~\bibnamefont {Yang}},\ }\bibfield  {title} {\bibinfo
  {title} {Measurement techniques for thermal conductivity and interfacial
  thermal conductance of bulk and thin film materials},\ }\href@noop {}
  {\bibfield  {journal} {\bibinfo  {journal} {J. Electron. Packag.}\ }\textbf
  {\bibinfo {volume} {138}},\ \bibinfo {pages} {040802} (\bibinfo {year}
  {2016})}\BibitemShut {NoStop}%
\bibitem [{\citenamefont {Majumdar}(1999)}]{majumdar1999scanning}%
  \BibitemOpen
  \bibfield  {author} {\bibinfo {author} {\bibfnamefont {A.}~\bibnamefont
  {Majumdar}},\ }\bibfield  {title} {\bibinfo {title} {Scanning thermal
  microscopy},\ }\href@noop {} {\bibfield  {journal} {\bibinfo  {journal}
  {Annu. Rev. Mater. Res.}\ }\textbf {\bibinfo {volume} {29}},\ \bibinfo
  {pages} {505} (\bibinfo {year} {1999})}\BibitemShut {NoStop}%
\bibitem [{\citenamefont {Gom{\`e}s}\ \emph {et~al.}(2015)\citenamefont
  {Gom{\`e}s}, \citenamefont {Assy},\ and\ \citenamefont
  {Chapuis}}]{gomes2015scanning}%
  \BibitemOpen
  \bibfield  {author} {\bibinfo {author} {\bibfnamefont {S.}~\bibnamefont
  {Gom{\`e}s}}, \bibinfo {author} {\bibfnamefont {A.}~\bibnamefont {Assy}},\
  and\ \bibinfo {author} {\bibfnamefont {P.-O.}\ \bibnamefont {Chapuis}},\
  }\bibfield  {title} {\bibinfo {title} {Scanning thermal microscopy: A
  review},\ }\href@noop {} {\bibfield  {journal} {\bibinfo  {journal} {Phys.
  Stat. Sol. A}\ }\textbf {\bibinfo {volume} {212}},\ \bibinfo {pages} {477}
  (\bibinfo {year} {2015})}\BibitemShut {NoStop}%
\bibitem [{\citenamefont {Zhang}\ \emph {et~al.}(2020)\citenamefont {Zhang},
  \citenamefont {Zhu}, \citenamefont {Hui}, \citenamefont {Lanza},
  \citenamefont {Borca-Tasciuc},\ and\ \citenamefont
  {Mu{\~n}oz~Rojo}}]{zhang2020review}%
  \BibitemOpen
  \bibfield  {author} {\bibinfo {author} {\bibfnamefont {Y.}~\bibnamefont
  {Zhang}}, \bibinfo {author} {\bibfnamefont {W.}~\bibnamefont {Zhu}}, \bibinfo
  {author} {\bibfnamefont {F.}~\bibnamefont {Hui}}, \bibinfo {author}
  {\bibfnamefont {M.}~\bibnamefont {Lanza}}, \bibinfo {author} {\bibfnamefont
  {T.}~\bibnamefont {Borca-Tasciuc}},\ and\ \bibinfo {author} {\bibfnamefont
  {M.}~\bibnamefont {Mu{\~n}oz~Rojo}},\ }\bibfield  {title} {\bibinfo {title}
  {A review on principles and applications of scanning thermal microscopy
  (sthm)},\ }\href@noop {} {\bibfield  {journal} {\bibinfo  {journal} {Adv.
  Funct. Mater.}\ }\textbf {\bibinfo {volume} {30}},\ \bibinfo {pages}
  {1900892} (\bibinfo {year} {2020})}\BibitemShut {NoStop}%
\bibitem [{\citenamefont {Williams}\ and\ \citenamefont
  {Wickramasinghe}(1986)}]{williams1986scanning}%
  \BibitemOpen
  \bibfield  {author} {\bibinfo {author} {\bibfnamefont {C.}~\bibnamefont
  {Williams}}\ and\ \bibinfo {author} {\bibfnamefont {H.}~\bibnamefont
  {Wickramasinghe}},\ }\bibfield  {title} {\bibinfo {title} {Scanning thermal
  profiler},\ }\href@noop {} {\bibfield  {journal} {\bibinfo  {journal} {Appl.
  Phys. Lett.}\ }\textbf {\bibinfo {volume} {49}},\ \bibinfo {pages} {1587}
  (\bibinfo {year} {1986})}\BibitemShut {NoStop}%
\bibitem [{\citenamefont {Weaver}\ \emph {et~al.}(1989)\citenamefont {Weaver},
  \citenamefont {Walpita},\ and\ \citenamefont
  {Wickramasinghe}}]{weaver1989optical}%
  \BibitemOpen
  \bibfield  {author} {\bibinfo {author} {\bibfnamefont {J.}~\bibnamefont
  {Weaver}}, \bibinfo {author} {\bibfnamefont {L.}~\bibnamefont {Walpita}},\
  and\ \bibinfo {author} {\bibfnamefont {H.}~\bibnamefont {Wickramasinghe}},\
  }\bibfield  {title} {\bibinfo {title} {Optical absorption microscopy and
  spectroscopy with nanometre resolution},\ }\href@noop {} {\bibfield
  {journal} {\bibinfo  {journal} {Nature}\ }\textbf {\bibinfo {volume} {342}},\
  \bibinfo {pages} {783} (\bibinfo {year} {1989})}\BibitemShut {NoStop}%
\bibitem [{\citenamefont {Zhou}\ \emph {et~al.}(1998)\citenamefont {Zhou},
  \citenamefont {Midha}, \citenamefont {Mills}, \citenamefont {Thoms},
  \citenamefont {Murad},\ and\ \citenamefont {Weaver}}]{zhou1998generic}%
  \BibitemOpen
  \bibfield  {author} {\bibinfo {author} {\bibfnamefont {H.}~\bibnamefont
  {Zhou}}, \bibinfo {author} {\bibfnamefont {A.}~\bibnamefont {Midha}},
  \bibinfo {author} {\bibfnamefont {G.}~\bibnamefont {Mills}}, \bibinfo
  {author} {\bibfnamefont {S.}~\bibnamefont {Thoms}}, \bibinfo {author}
  {\bibfnamefont {S.}~\bibnamefont {Murad}},\ and\ \bibinfo {author}
  {\bibfnamefont {J.}~\bibnamefont {Weaver}},\ }\bibfield  {title} {\bibinfo
  {title} {Generic scanned-probe microscope sensors by combined micromachining
  and electron-beam lithography},\ }\href@noop {} {\bibfield  {journal}
  {\bibinfo  {journal} {J. Vac. Sci. Techno. B}\ }\textbf {\bibinfo {volume}
  {16}},\ \bibinfo {pages} {54} (\bibinfo {year} {1998})}\BibitemShut {NoStop}%
\bibitem [{\citenamefont {Majumdar}\ \emph {et~al.}(1993)\citenamefont
  {Majumdar}, \citenamefont {Carrejo},\ and\ \citenamefont
  {Lai}}]{majumdar1993thermal}%
  \BibitemOpen
  \bibfield  {author} {\bibinfo {author} {\bibfnamefont {A.}~\bibnamefont
  {Majumdar}}, \bibinfo {author} {\bibfnamefont {J.}~\bibnamefont {Carrejo}},\
  and\ \bibinfo {author} {\bibfnamefont {J.}~\bibnamefont {Lai}},\ }\bibfield
  {title} {\bibinfo {title} {Thermal imaging using the atomic force
  microscope},\ }\href@noop {} {\bibfield  {journal} {\bibinfo  {journal}
  {Appl. Phys. Lett.}\ }\textbf {\bibinfo {volume} {62}},\ \bibinfo {pages}
  {2501} (\bibinfo {year} {1993})}\BibitemShut {NoStop}%
\bibitem [{\citenamefont {Shi}\ \emph {et~al.}(2001)\citenamefont {Shi},
  \citenamefont {Kwon}, \citenamefont {Miner},\ and\ \citenamefont
  {Majumdar}}]{shi2001design}%
  \BibitemOpen
  \bibfield  {author} {\bibinfo {author} {\bibfnamefont {L.}~\bibnamefont
  {Shi}}, \bibinfo {author} {\bibfnamefont {O.}~\bibnamefont {Kwon}}, \bibinfo
  {author} {\bibfnamefont {A.~C.}\ \bibnamefont {Miner}},\ and\ \bibinfo
  {author} {\bibfnamefont {A.}~\bibnamefont {Majumdar}},\ }\bibfield  {title}
  {\bibinfo {title} {Design and batch fabrication of probes for sub-100 nm
  scanning thermal microscopy},\ }\href@noop {} {\bibfield  {journal} {\bibinfo
   {journal} {J. Microelectron. Syst.}\ }\textbf {\bibinfo {volume} {10}},\
  \bibinfo {pages} {370} (\bibinfo {year} {2001})}\BibitemShut {NoStop}%
\bibitem [{\citenamefont {Kim}\ \emph {et~al.}(2012)\citenamefont {Kim},
  \citenamefont {Jeong}, \citenamefont {Lee},\ and\ \citenamefont
  {Reddy}}]{kim2012ultra}%
  \BibitemOpen
  \bibfield  {author} {\bibinfo {author} {\bibfnamefont {K.}~\bibnamefont
  {Kim}}, \bibinfo {author} {\bibfnamefont {W.}~\bibnamefont {Jeong}}, \bibinfo
  {author} {\bibfnamefont {W.}~\bibnamefont {Lee}},\ and\ \bibinfo {author}
  {\bibfnamefont {P.}~\bibnamefont {Reddy}},\ }\bibfield  {title} {\bibinfo
  {title} {Ultra-high vacuum scanning thermal microscopy for nanometer
  resolution quantitative thermometry},\ }\href@noop {} {\bibfield  {journal}
  {\bibinfo  {journal} {ACS Nano}\ }\textbf {\bibinfo {volume} {6}},\ \bibinfo
  {pages} {4248} (\bibinfo {year} {2012})}\BibitemShut {NoStop}%
\bibitem [{\citenamefont {Nguyen}\ \emph
  {et~al.}(2019{\natexlab{a}})\citenamefont {Nguyen}, \citenamefont {Thiery},
  \citenamefont {Euphrasie}, \citenamefont {Gomes}, \citenamefont {Hay},\ and\
  \citenamefont {Vairac}}]{nguyen2019}%
  \BibitemOpen
  \bibfield  {author} {\bibinfo {author} {\bibfnamefont {T.~P.}\ \bibnamefont
  {Nguyen}}, \bibinfo {author} {\bibfnamefont {L.}~\bibnamefont {Thiery}},
  \bibinfo {author} {\bibfnamefont {S.}~\bibnamefont {Euphrasie}}, \bibinfo
  {author} {\bibfnamefont {S.}~\bibnamefont {Gomes}}, \bibinfo {author}
  {\bibfnamefont {B.}~\bibnamefont {Hay}},\ and\ \bibinfo {author}
  {\bibfnamefont {P.}~\bibnamefont {Vairac}},\ }\bibfield  {title} {\bibinfo
  {title} {Calibration of thermocouple-based scanning thermal microscope in
  active mode (2-omega method)},\ }\href {https://doi.org/10.1063/1.5119044}
  {\bibfield  {journal} {\bibinfo  {journal} {Rev. Sci. Instrum.}\ }\textbf
  {\bibinfo {volume} {90}},\ \bibinfo {pages} {114901} (\bibinfo {year}
  {2019}{\natexlab{a}})}\BibitemShut {NoStop}%
\bibitem [{\citenamefont {Pylkki}\ \emph {et~al.}(1994)\citenamefont {Pylkki},
  \citenamefont {Moyer},\ and\ \citenamefont {West}}]{pylkki1994scanning}%
  \BibitemOpen
  \bibfield  {author} {\bibinfo {author} {\bibfnamefont {R.~J.}\ \bibnamefont
  {Pylkki}}, \bibinfo {author} {\bibfnamefont {P.~J. M. P.~J.}\ \bibnamefont
  {Moyer}},\ and\ \bibinfo {author} {\bibfnamefont {P.~E. W. P.~E.}\
  \bibnamefont {West}},\ }\bibfield  {title} {\bibinfo {title} {Scanning
  near-field optical microscopy and scanning thermal microscopy},\ }\href@noop
  {} {\bibfield  {journal} {\bibinfo  {journal} {Jpn. J. Appl. Phys.}\ }\textbf
  {\bibinfo {volume} {33}},\ \bibinfo {pages} {3785} (\bibinfo {year}
  {1994})}\BibitemShut {NoStop}%
\bibitem [{\citenamefont {Lef{\`e}vre}\ and\ \citenamefont
  {Volz}(2005)}]{lefevre20053}%
  \BibitemOpen
  \bibfield  {author} {\bibinfo {author} {\bibfnamefont {S.}~\bibnamefont
  {Lef{\`e}vre}}\ and\ \bibinfo {author} {\bibfnamefont {S.}~\bibnamefont
  {Volz}},\ }\bibfield  {title} {\bibinfo {title} {3 omega-scanning thermal
  microscope},\ }\href@noop {} {\bibfield  {journal} {\bibinfo  {journal} {Rev.
  Sci. Instrum.}\ }\textbf {\bibinfo {volume} {76}},\ \bibinfo {pages} {033701}
  (\bibinfo {year} {2005})}\BibitemShut {NoStop}%
\bibitem [{\citenamefont {Janus}\ \emph {et~al.}(2018)\citenamefont {Janus},
  \citenamefont {Sierakowski}, \citenamefont {Rudek}, \citenamefont {Kunicki},
  \citenamefont {Dzierka}, \citenamefont {Biczysko},\ and\ \citenamefont
  {Gotszalk}}]{janus2018}%
  \BibitemOpen
  \bibfield  {author} {\bibinfo {author} {\bibfnamefont {P.}~\bibnamefont
  {Janus}}, \bibinfo {author} {\bibfnamefont {A.}~\bibnamefont {Sierakowski}},
  \bibinfo {author} {\bibfnamefont {M.}~\bibnamefont {Rudek}}, \bibinfo
  {author} {\bibfnamefont {P.}~\bibnamefont {Kunicki}}, \bibinfo {author}
  {\bibfnamefont {A.}~\bibnamefont {Dzierka}}, \bibinfo {author} {\bibfnamefont
  {P.}~\bibnamefont {Biczysko}},\ and\ \bibinfo {author} {\bibfnamefont
  {T.}~\bibnamefont {Gotszalk}},\ }\bibfield  {title} {\bibinfo {title}
  {Thermal nanometrology using piezoresistive sthm probes with metallic tips},\
  }\href {https://doi.org/10.1016/j.ultramic.2018.06.016} {\bibfield  {journal}
  {\bibinfo  {journal} {Ultramicroscopy}\ }\textbf {\bibinfo {volume} {193}},\
  \bibinfo {pages} {104} (\bibinfo {year} {2018})}\BibitemShut {NoStop}%
\bibitem [{\citenamefont {Bodzenta}\ \emph {et~al.}(2020)\citenamefont
  {Bodzenta}, \citenamefont {Ka{\'z}mierczak-Ba{\l}ata},\ and\ \citenamefont
  {Harris}}]{bodzenta2020}%
  \BibitemOpen
  \bibfield  {author} {\bibinfo {author} {\bibfnamefont {J.}~\bibnamefont
  {Bodzenta}}, \bibinfo {author} {\bibfnamefont {A.}~\bibnamefont
  {Ka{\'z}mierczak-Ba{\l}ata}},\ and\ \bibinfo {author} {\bibfnamefont
  {K.}~\bibnamefont {Harris}},\ }\bibfield  {title} {\bibinfo {title}
  {Quantitative thermal measurement by the use of scanning thermal microscope
  and resistive thermal probes},\ }\href@noop {} {\bibfield  {journal}
  {\bibinfo  {journal} {J. Appl. Phys.}\ }\textbf {\bibinfo {volume} {127}},\
  \bibinfo {pages} {031103} (\bibinfo {year} {2020})}\BibitemShut {NoStop}%
\bibitem [{\citenamefont {Pernot}\ \emph {et~al.}(2021)\citenamefont {Pernot},
  \citenamefont {Metjari}, \citenamefont {Chaynes}, \citenamefont {Weber},
  \citenamefont {Isaiev},\ and\ \citenamefont {Lacroix}}]{pernot2021frequency}%
  \BibitemOpen
  \bibfield  {author} {\bibinfo {author} {\bibfnamefont {G.}~\bibnamefont
  {Pernot}}, \bibinfo {author} {\bibfnamefont {A.}~\bibnamefont {Metjari}},
  \bibinfo {author} {\bibfnamefont {H.}~\bibnamefont {Chaynes}}, \bibinfo
  {author} {\bibfnamefont {M.}~\bibnamefont {Weber}}, \bibinfo {author}
  {\bibfnamefont {M.}~\bibnamefont {Isaiev}},\ and\ \bibinfo {author}
  {\bibfnamefont {D.}~\bibnamefont {Lacroix}},\ }\bibfield  {title} {\bibinfo
  {title} {Frequency domain analysis of 3$\omega$-scanning thermal microscope
  probe—application to tip/surface thermal interface measurements in vacuum
  environment},\ }\href@noop {} {\bibfield  {journal} {\bibinfo  {journal} {J.
  Appl. Phys}\ }\textbf {\bibinfo {volume} {129}},\ \bibinfo {pages} {055105}
  (\bibinfo {year} {2021})}\BibitemShut {NoStop}%
\bibitem [{\citenamefont {Nakabeppu}\ \emph {et~al.}(1995)\citenamefont
  {Nakabeppu}, \citenamefont {Chandrachood}, \citenamefont {Wu}, \citenamefont
  {Lai},\ and\ \citenamefont {Majumdar}}]{nakabeppu1995scanning}%
  \BibitemOpen
  \bibfield  {author} {\bibinfo {author} {\bibfnamefont {O.}~\bibnamefont
  {Nakabeppu}}, \bibinfo {author} {\bibfnamefont {M.}~\bibnamefont
  {Chandrachood}}, \bibinfo {author} {\bibfnamefont {Y.}~\bibnamefont {Wu}},
  \bibinfo {author} {\bibfnamefont {J.}~\bibnamefont {Lai}},\ and\ \bibinfo
  {author} {\bibfnamefont {A.}~\bibnamefont {Majumdar}},\ }\bibfield  {title}
  {\bibinfo {title} {Scanning thermal imaging microscopy using composite
  cantilever probes},\ }\href@noop {} {\bibfield  {journal} {\bibinfo
  {journal} {Appl. Phys. Lett.}\ }\textbf {\bibinfo {volume} {66}},\ \bibinfo
  {pages} {694} (\bibinfo {year} {1995})}\BibitemShut {NoStop}%
\bibitem [{\citenamefont {Samson}\ \emph {et~al.}(2008)\citenamefont {Samson},
  \citenamefont {Aigouy}, \citenamefont {L{\"o}w}, \citenamefont {Bergaud},
  \citenamefont {Kim},\ and\ \citenamefont {Mortier}}]{samson2008ac}%
  \BibitemOpen
  \bibfield  {author} {\bibinfo {author} {\bibfnamefont {B.}~\bibnamefont
  {Samson}}, \bibinfo {author} {\bibfnamefont {L.}~\bibnamefont {Aigouy}},
  \bibinfo {author} {\bibfnamefont {P.}~\bibnamefont {L{\"o}w}}, \bibinfo
  {author} {\bibfnamefont {C.}~\bibnamefont {Bergaud}}, \bibinfo {author}
  {\bibfnamefont {B.}~\bibnamefont {Kim}},\ and\ \bibinfo {author}
  {\bibfnamefont {M.}~\bibnamefont {Mortier}},\ }\bibfield  {title} {\bibinfo
  {title} {ac thermal imaging of nanoheaters using a scanning fluorescent
  probe},\ }\href@noop {} {\bibfield  {journal} {\bibinfo  {journal} {Appl.
  Phys. Lett.}\ }\textbf {\bibinfo {volume} {92}},\ \bibinfo {pages} {023101}
  (\bibinfo {year} {2008})}\BibitemShut {NoStop}%
\bibitem [{\citenamefont {Hatakeyama}\ \emph {et~al.}(2014)\citenamefont
  {Hatakeyama}, \citenamefont {Sarajlic}, \citenamefont {Siekman},
  \citenamefont {Jalabert}, \citenamefont {Fujita}, \citenamefont {Tas},\ and\
  \citenamefont {Abelmann}}]{hatakeyama2014wafer}%
  \BibitemOpen
  \bibfield  {author} {\bibinfo {author} {\bibfnamefont {K.}~\bibnamefont
  {Hatakeyama}}, \bibinfo {author} {\bibfnamefont {E.}~\bibnamefont
  {Sarajlic}}, \bibinfo {author} {\bibfnamefont {M.~H.}\ \bibnamefont
  {Siekman}}, \bibinfo {author} {\bibfnamefont {L.}~\bibnamefont {Jalabert}},
  \bibinfo {author} {\bibfnamefont {H.}~\bibnamefont {Fujita}}, \bibinfo
  {author} {\bibfnamefont {N.}~\bibnamefont {Tas}},\ and\ \bibinfo {author}
  {\bibfnamefont {L.}~\bibnamefont {Abelmann}},\ }\bibfield  {title} {\bibinfo
  {title} {Wafer-scale fabrication of scanning thermal probes with integrated
  metal nanowire resistive elements for sensing and heating},\ }in\ \href@noop
  {} {\emph {\bibinfo {booktitle} {Int. Conf. on Micro Electro Mech.
  Systems}}}\ (\bibinfo {organization} {IEEE},\ \bibinfo {year} {2014})\ p.\
  \bibinfo {pages} {1111}\BibitemShut {NoStop}%
\bibitem [{\citenamefont {Dobson}\ \emph {et~al.}(2007)\citenamefont {Dobson},
  \citenamefont {Weaver},\ and\ \citenamefont {Mills}}]{dobson2007new}%
  \BibitemOpen
  \bibfield  {author} {\bibinfo {author} {\bibfnamefont {P.~S.}\ \bibnamefont
  {Dobson}}, \bibinfo {author} {\bibfnamefont {J.~M.}\ \bibnamefont {Weaver}},\
  and\ \bibinfo {author} {\bibfnamefont {G.}~\bibnamefont {Mills}},\ }\bibfield
   {title} {\bibinfo {title} {New methods for calibrated scanning thermal
  microscopy (sthm)},\ }in\ \href@noop {} {\emph {\bibinfo {booktitle} {Proc.
  IEEE Sens.}}}\ (\bibinfo {organization} {IEEE},\ \bibinfo {year} {2007})\ p.\
  \bibinfo {pages} {708}\BibitemShut {NoStop}%
\bibitem [{\citenamefont {Puyoo}\ \emph {et~al.}(2010)\citenamefont {Puyoo},
  \citenamefont {Grauby}, \citenamefont {Rampnoux}, \citenamefont
  {Rouvi{\`e}re},\ and\ \citenamefont {Dilhaire}}]{puyoo2010thermal}%
  \BibitemOpen
  \bibfield  {author} {\bibinfo {author} {\bibfnamefont {E.}~\bibnamefont
  {Puyoo}}, \bibinfo {author} {\bibfnamefont {S.}~\bibnamefont {Grauby}},
  \bibinfo {author} {\bibfnamefont {J.-M.}\ \bibnamefont {Rampnoux}}, \bibinfo
  {author} {\bibfnamefont {E.}~\bibnamefont {Rouvi{\`e}re}},\ and\ \bibinfo
  {author} {\bibfnamefont {S.}~\bibnamefont {Dilhaire}},\ }\bibfield  {title}
  {\bibinfo {title} {Thermal exchange radius measurement: Application to
  nanowire thermal imaging},\ }\href@noop {} {\bibfield  {journal} {\bibinfo
  {journal} {Rev. Sci. Instrum.}\ }\textbf {\bibinfo {volume} {81}},\ \bibinfo
  {pages} {073701} (\bibinfo {year} {2010})}\BibitemShut {NoStop}%
\bibitem [{\citenamefont {Rangelow}\ \emph {et~al.}(2001)\citenamefont
  {Rangelow}, \citenamefont {Gotszalk}, \citenamefont {Grabiec}, \citenamefont
  {Edinger},\ and\ \citenamefont {Abedinov}}]{rangelow2001thermal}%
  \BibitemOpen
  \bibfield  {author} {\bibinfo {author} {\bibfnamefont {I.}~\bibnamefont
  {Rangelow}}, \bibinfo {author} {\bibfnamefont {T.}~\bibnamefont {Gotszalk}},
  \bibinfo {author} {\bibfnamefont {P.}~\bibnamefont {Grabiec}}, \bibinfo
  {author} {\bibfnamefont {K.}~\bibnamefont {Edinger}},\ and\ \bibinfo {author}
  {\bibfnamefont {N.}~\bibnamefont {Abedinov}},\ }\bibfield  {title} {\bibinfo
  {title} {Thermal nano-probe},\ }\href@noop {} {\bibfield  {journal} {\bibinfo
   {journal} {Microelectron. Eng.}\ }\textbf {\bibinfo {volume} {57}},\
  \bibinfo {pages} {737} (\bibinfo {year} {2001})}\BibitemShut {NoStop}%
\bibitem [{\citenamefont {Lefevre}\ \emph {et~al.}(2004)\citenamefont
  {Lefevre}, \citenamefont {Saulnier}, \citenamefont {Fuentes},\ and\
  \citenamefont {Volz}}]{lefevre2004probe}%
  \BibitemOpen
  \bibfield  {author} {\bibinfo {author} {\bibfnamefont {S.}~\bibnamefont
  {Lefevre}}, \bibinfo {author} {\bibfnamefont {J.-B.}\ \bibnamefont
  {Saulnier}}, \bibinfo {author} {\bibfnamefont {C.}~\bibnamefont {Fuentes}},\
  and\ \bibinfo {author} {\bibfnamefont {S.}~\bibnamefont {Volz}},\ }\bibfield
  {title} {\bibinfo {title} {Probe calibration of the scanning thermal
  microscope in the ac mode},\ }\href@noop {} {\bibfield  {journal} {\bibinfo
  {journal} {Superlattices and Microstruct.}\ }\textbf {\bibinfo {volume}
  {35}},\ \bibinfo {pages} {283} (\bibinfo {year} {2004})}\BibitemShut
  {NoStop}%
\bibitem [{\citenamefont {Nelson}\ and\ \citenamefont
  {King}(2007)}]{nelson2007measuring}%
  \BibitemOpen
  \bibfield  {author} {\bibinfo {author} {\bibfnamefont {B.~A.}\ \bibnamefont
  {Nelson}}\ and\ \bibinfo {author} {\bibfnamefont {W.}~\bibnamefont {King}},\
  }\bibfield  {title} {\bibinfo {title} {Measuring material softening with
  nanoscale spatial resolution using heated silicon probes},\ }\href@noop {}
  {\bibfield  {journal} {\bibinfo  {journal} {Rev. Sci. Instrum.}\ }\textbf
  {\bibinfo {volume} {78}},\ \bibinfo {pages} {023702} (\bibinfo {year}
  {2007})}\BibitemShut {NoStop}%
\bibitem [{\citenamefont {Swami}\ \emph {et~al.}(2022)\citenamefont {Swami},
  \citenamefont {Julie}, \citenamefont {Singhal}, \citenamefont {Paterson},
  \citenamefont {Maire}, \citenamefont {Le-Denmat}, \citenamefont {Motte},
  \citenamefont {Gomes},\ and\ \citenamefont {Bourgeois}}]{swami2022electron}%
  \BibitemOpen
  \bibfield  {author} {\bibinfo {author} {\bibfnamefont {R.}~\bibnamefont
  {Swami}}, \bibinfo {author} {\bibfnamefont {G.}~\bibnamefont {Julie}},
  \bibinfo {author} {\bibfnamefont {D.}~\bibnamefont {Singhal}}, \bibinfo
  {author} {\bibfnamefont {J.}~\bibnamefont {Paterson}}, \bibinfo {author}
  {\bibfnamefont {J.}~\bibnamefont {Maire}}, \bibinfo {author} {\bibfnamefont
  {S.}~\bibnamefont {Le-Denmat}}, \bibinfo {author} {\bibfnamefont {J.~F.}\
  \bibnamefont {Motte}}, \bibinfo {author} {\bibfnamefont {S.}~\bibnamefont
  {Gomes}},\ and\ \bibinfo {author} {\bibfnamefont {O.}~\bibnamefont
  {Bourgeois}},\ }\bibfield  {title} {\bibinfo {title} {Electron beam
  lithography on non-planar, suspended, 3d afm cantilever for nanoscale thermal
  probing},\ }\href@noop {} {\bibfield  {journal} {\bibinfo  {journal} {Nano
  Futures}\ }\textbf {\bibinfo {volume} {6}},\ \bibinfo {pages} {025005}
  (\bibinfo {year} {2022})}\BibitemShut {NoStop}%
\bibitem [{\citenamefont {Chirtoc}\ \emph {et~al.}(2008)\citenamefont
  {Chirtoc}, \citenamefont {Gibkes}, \citenamefont {Wernhardt}, \citenamefont
  {Pelzl},\ and\ \citenamefont {Wieck}}]{chirtoc2008}%
  \BibitemOpen
  \bibfield  {author} {\bibinfo {author} {\bibfnamefont {M.}~\bibnamefont
  {Chirtoc}}, \bibinfo {author} {\bibfnamefont {J.}~\bibnamefont {Gibkes}},
  \bibinfo {author} {\bibfnamefont {R.}~\bibnamefont {Wernhardt}}, \bibinfo
  {author} {\bibfnamefont {J.}~\bibnamefont {Pelzl}},\ and\ \bibinfo {author}
  {\bibfnamefont {A.}~\bibnamefont {Wieck}},\ }\bibfield  {title} {\bibinfo
  {title} {Temperature-dependent quantitative 3-omega scanning thermal
  microscopy: Local thermal conductivity changes in niti microstructures
  induced by martensite-austenite phase transition},\ }\href
  {https://doi.org/10.1063/1.2982235} {\bibfield  {journal} {\bibinfo
  {journal} {Rev. Sci. Instrum.}\ }\textbf {\bibinfo {volume} {79}},\ \bibinfo
  {pages} {093703} (\bibinfo {year} {2008})}\BibitemShut {NoStop}%
\bibitem [{\citenamefont {Bodzenta}\ \emph {et~al.}(2016)\citenamefont
  {Bodzenta}, \citenamefont {Juszczyk}, \citenamefont
  {Ka{\'z}mierczak-Ba{\l}ata}, \citenamefont {Firek}, \citenamefont {Fleming},\
  and\ \citenamefont {Chirtoc}}]{bodzenta2016}%
  \BibitemOpen
  \bibfield  {author} {\bibinfo {author} {\bibfnamefont {J.}~\bibnamefont
  {Bodzenta}}, \bibinfo {author} {\bibfnamefont {J.}~\bibnamefont {Juszczyk}},
  \bibinfo {author} {\bibfnamefont {A.}~\bibnamefont
  {Ka{\'z}mierczak-Ba{\l}ata}}, \bibinfo {author} {\bibfnamefont
  {P.}~\bibnamefont {Firek}}, \bibinfo {author} {\bibfnamefont
  {A.}~\bibnamefont {Fleming}},\ and\ \bibinfo {author} {\bibfnamefont
  {M.}~\bibnamefont {Chirtoc}},\ }\bibfield  {title} {\bibinfo {title}
  {Quantitative thermal microscopy measurement with thermal probe driven by dc+
  ac current},\ }\href@noop {} {\bibfield  {journal} {\bibinfo  {journal} {Int.
  J. Thermophys}\ }\textbf {\bibinfo {volume} {37}},\ \bibinfo {pages} {1}
  (\bibinfo {year} {2016})}\BibitemShut {NoStop}%
\bibitem [{\citenamefont {Assy}\ and\ \citenamefont
  {Gomes}(2015{\natexlab{a}})}]{AssyAPL2015}%
  \BibitemOpen
  \bibfield  {author} {\bibinfo {author} {\bibfnamefont {A.}~\bibnamefont
  {Assy}}\ and\ \bibinfo {author} {\bibfnamefont {S.}~\bibnamefont {Gomes}},\
  }\bibfield  {title} {\bibinfo {title} {Heat transfer at nanoscale contacts
  investigated with scanning thermal microscopy},\ }\bibfield  {journal}
  {\bibinfo  {journal} {Appl. Phys. Lett.}\ }\textbf {\bibinfo {volume}
  {107}},\ \href {https://doi.org/10.1063/1.4927653} {10.1063/1.4927653}
  (\bibinfo {year} {2015}{\natexlab{a}})\BibitemShut {NoStop}%
\bibitem [{\citenamefont {Assy}\ and\ \citenamefont
  {Gomes}(2015{\natexlab{b}})}]{AssyNanotech2015}%
  \BibitemOpen
  \bibfield  {author} {\bibinfo {author} {\bibfnamefont {A.}~\bibnamefont
  {Assy}}\ and\ \bibinfo {author} {\bibfnamefont {S.}~\bibnamefont {Gomes}},\
  }\bibfield  {title} {\bibinfo {title} {Temperature-dependent capillary forces
  at nano-contacts for estimating the heat conduction through a water
  meniscus},\ }\href {https://doi.org/10.1088/0957-4484/26/35/355401}
  {\bibfield  {journal} {\bibinfo  {journal} {Nanotechnology}\ }\textbf
  {\bibinfo {volume} {26}},\ \bibinfo {pages} {477} (\bibinfo {year}
  {2015}{\natexlab{b}})}\BibitemShut {NoStop}%
\bibitem [{\citenamefont {Gotsmann}\ and\ \citenamefont
  {Lantz}(2013)}]{gotsmann}%
  \BibitemOpen
  \bibfield  {author} {\bibinfo {author} {\bibfnamefont {B.}~\bibnamefont
  {Gotsmann}}\ and\ \bibinfo {author} {\bibfnamefont {M.}~\bibnamefont
  {Lantz}},\ }\bibfield  {title} {\bibinfo {title} {Quantized thermal transport
  across contacts of rough surfaces},\ }\href
  {https://doi.org/10.1038/NMAT3460} {\bibfield  {journal} {\bibinfo  {journal}
  {Nat. Materials}\ }\textbf {\bibinfo {volume} {12}},\ \bibinfo {pages} {59}
  (\bibinfo {year} {2013})}\BibitemShut {NoStop}%
\bibitem [{\citenamefont {Guen}\ \emph {et~al.}(2021)\citenamefont {Guen},
  \citenamefont {Chapuis}, \citenamefont {Kaur}, \citenamefont {Klapetek},\
  and\ \citenamefont {Gomes}}]{guen}%
  \BibitemOpen
  \bibfield  {author} {\bibinfo {author} {\bibfnamefont {E.}~\bibnamefont
  {Guen}}, \bibinfo {author} {\bibfnamefont {P.-O.}\ \bibnamefont {Chapuis}},
  \bibinfo {author} {\bibfnamefont {N.~J.}\ \bibnamefont {Kaur}}, \bibinfo
  {author} {\bibfnamefont {P.}~\bibnamefont {Klapetek}},\ and\ \bibinfo
  {author} {\bibfnamefont {S.}~\bibnamefont {Gomes}},\ }\bibfield  {title}
  {\bibinfo {title} {Impact of roughness on heat conduction involving
  nanocontacts},\ }\bibfield  {journal} {\bibinfo  {journal} {Appl. Phys.
  Lett.}\ }\textbf {\bibinfo {volume} {119}},\ \href
  {https://doi.org/10.1063/5.0064244} {10.1063/5.0064244} (\bibinfo {year}
  {2021})\BibitemShut {NoStop}%
\bibitem [{\citenamefont {Cahill}(1990)}]{cahill1990thermal}%
  \BibitemOpen
  \bibfield  {author} {\bibinfo {author} {\bibfnamefont {D.~G.}\ \bibnamefont
  {Cahill}},\ }\bibfield  {title} {\bibinfo {title} {Thermal conductivity
  measurement from 30 to 750 k: the 3-omega method},\ }\href@noop {} {\bibfield
   {journal} {\bibinfo  {journal} {Rev. Sci. Instrum.}\ }\textbf {\bibinfo
  {volume} {61}},\ \bibinfo {pages} {802} (\bibinfo {year} {1990})}\BibitemShut
  {NoStop}%
\bibitem [{\citenamefont {Sandell}\ \emph {et~al.}(2020)\citenamefont
  {Sandell}, \citenamefont {Ch{\'a}vez-{\'A}ngel}, \citenamefont {El~Sachat},
  \citenamefont {He}, \citenamefont {Sotomayor~Torres},\ and\ \citenamefont
  {Maire}}]{sandell2020thermoreflectance}%
  \BibitemOpen
  \bibfield  {author} {\bibinfo {author} {\bibfnamefont {S.}~\bibnamefont
  {Sandell}}, \bibinfo {author} {\bibfnamefont {E.}~\bibnamefont
  {Ch{\'a}vez-{\'A}ngel}}, \bibinfo {author} {\bibfnamefont {A.}~\bibnamefont
  {El~Sachat}}, \bibinfo {author} {\bibfnamefont {J.}~\bibnamefont {He}},
  \bibinfo {author} {\bibfnamefont {C.~M.}\ \bibnamefont {Sotomayor~Torres}},\
  and\ \bibinfo {author} {\bibfnamefont {J.}~\bibnamefont {Maire}},\ }\bibfield
   {title} {\bibinfo {title} {Thermoreflectance techniques and raman
  thermometry for thermal property characterization of nanostructures},\
  }\href@noop {} {\bibfield  {journal} {\bibinfo  {journal} {J. Appl. Phys.}\
  }\textbf {\bibinfo {volume} {128}},\ \bibinfo {pages} {131101} (\bibinfo
  {year} {2020})}\BibitemShut {NoStop}%
\bibitem [{\citenamefont {Bourgeois}\ \emph {et~al.}(2006)\citenamefont
  {Bourgeois}, \citenamefont {Andr{\'e}}, \citenamefont {Macovei},\ and\
  \citenamefont {Chaussy}}]{bourgeois2006liquid}%
  \BibitemOpen
  \bibfield  {author} {\bibinfo {author} {\bibfnamefont {O.}~\bibnamefont
  {Bourgeois}}, \bibinfo {author} {\bibfnamefont {E.}~\bibnamefont
  {Andr{\'e}}}, \bibinfo {author} {\bibfnamefont {C.}~\bibnamefont {Macovei}},\
  and\ \bibinfo {author} {\bibfnamefont {J.}~\bibnamefont {Chaussy}},\
  }\bibfield  {title} {\bibinfo {title} {Liquid nitrogen to room-temperature
  thermometry using niobium nitride thin films},\ }\href@noop {} {\bibfield
  {journal} {\bibinfo  {journal} {Rev. Sci. Instrum.}\ }\textbf {\bibinfo
  {volume} {77}},\ \bibinfo {pages} {126108} (\bibinfo {year}
  {2006})}\BibitemShut {NoStop}%
\bibitem [{\citenamefont {Nguyen}\ \emph
  {et~al.}(2019{\natexlab{b}})\citenamefont {Nguyen}, \citenamefont {Tavakoli},
  \citenamefont {Triqueneaux}, \citenamefont {Swami}, \citenamefont {Ruhtinas},
  \citenamefont {Gradel}, \citenamefont {Garcia-Campos}, \citenamefont
  {Hasselbach}, \citenamefont {Frydman}, \citenamefont {Piot} \emph
  {et~al.}}]{nguyen2019niobium}%
  \BibitemOpen
  \bibfield  {author} {\bibinfo {author} {\bibfnamefont {T.}~\bibnamefont
  {Nguyen}}, \bibinfo {author} {\bibfnamefont {A.}~\bibnamefont {Tavakoli}},
  \bibinfo {author} {\bibfnamefont {S.}~\bibnamefont {Triqueneaux}}, \bibinfo
  {author} {\bibfnamefont {R.}~\bibnamefont {Swami}}, \bibinfo {author}
  {\bibfnamefont {A.}~\bibnamefont {Ruhtinas}}, \bibinfo {author}
  {\bibfnamefont {J.}~\bibnamefont {Gradel}}, \bibinfo {author} {\bibfnamefont
  {P.}~\bibnamefont {Garcia-Campos}}, \bibinfo {author} {\bibfnamefont
  {K.}~\bibnamefont {Hasselbach}}, \bibinfo {author} {\bibfnamefont
  {A.}~\bibnamefont {Frydman}}, \bibinfo {author} {\bibfnamefont
  {B.}~\bibnamefont {Piot}}, \emph {et~al.},\ }\bibfield  {title} {\bibinfo
  {title} {Niobium nitride thin films for very low temperature resistive
  thermometry},\ }\href@noop {} {\bibfield  {journal} {\bibinfo  {journal} {J.
  Low Temp. Phys.}\ }\textbf {\bibinfo {volume} {197}},\ \bibinfo {pages} {348}
  (\bibinfo {year} {2019}{\natexlab{b}})}\BibitemShut {NoStop}%
\bibitem [{\citenamefont {Ftouni}\ \emph {et~al.}(2015)\citenamefont {Ftouni},
  \citenamefont {Blanc}, \citenamefont {Tainoff}, \citenamefont {Fefferman},
  \citenamefont {Defoort}, \citenamefont {Lulla}, \citenamefont {Richard},
  \citenamefont {Collin},\ and\ \citenamefont {Bourgeois}}]{ftouni2015thermal}%
  \BibitemOpen
  \bibfield  {author} {\bibinfo {author} {\bibfnamefont {H.}~\bibnamefont
  {Ftouni}}, \bibinfo {author} {\bibfnamefont {C.}~\bibnamefont {Blanc}},
  \bibinfo {author} {\bibfnamefont {D.}~\bibnamefont {Tainoff}}, \bibinfo
  {author} {\bibfnamefont {A.~D.}\ \bibnamefont {Fefferman}}, \bibinfo {author}
  {\bibfnamefont {M.}~\bibnamefont {Defoort}}, \bibinfo {author} {\bibfnamefont
  {K.~J.}\ \bibnamefont {Lulla}}, \bibinfo {author} {\bibfnamefont
  {J.}~\bibnamefont {Richard}}, \bibinfo {author} {\bibfnamefont
  {E.}~\bibnamefont {Collin}},\ and\ \bibinfo {author} {\bibfnamefont
  {O.}~\bibnamefont {Bourgeois}},\ }\bibfield  {title} {\bibinfo {title}
  {Thermal conductivity of silicon nitride membranes is not sensitive to
  stress},\ }\href@noop {} {\bibfield  {journal} {\bibinfo  {journal} {Phys.
  Rev. B}\ }\textbf {\bibinfo {volume} {92}},\ \bibinfo {pages} {125439}
  (\bibinfo {year} {2015})}\BibitemShut {NoStop}%
\bibitem [{\citenamefont {Heron}\ \emph {et~al.}(2009)\citenamefont {Heron},
  \citenamefont {Fournier}, \citenamefont {Mingo},\ and\ \citenamefont
  {Bourgeois}}]{heron2009mesoscopic}%
  \BibitemOpen
  \bibfield  {author} {\bibinfo {author} {\bibfnamefont {J.}~\bibnamefont
  {Heron}}, \bibinfo {author} {\bibfnamefont {T.}~\bibnamefont {Fournier}},
  \bibinfo {author} {\bibfnamefont {N.}~\bibnamefont {Mingo}},\ and\ \bibinfo
  {author} {\bibfnamefont {O.}~\bibnamefont {Bourgeois}},\ }\bibfield  {title}
  {\bibinfo {title} {Mesoscopic size effects on the thermal conductance of
  silicon nanowire},\ }\href@noop {} {\bibfield  {journal} {\bibinfo  {journal}
  {Nano Lett.}\ }\textbf {\bibinfo {volume} {9}},\ \bibinfo {pages} {1861}
  (\bibinfo {year} {2009})}\BibitemShut {NoStop}%
\bibitem [{\citenamefont {Fiege}\ \emph {et~al.}(1999)\citenamefont {Fiege},
  \citenamefont {Altes}, \citenamefont {Heiderhoff},\ and\ \citenamefont
  {Balk}}]{fiege1999}%
  \BibitemOpen
  \bibfield  {author} {\bibinfo {author} {\bibfnamefont {G.~B.~M.}\
  \bibnamefont {Fiege}}, \bibinfo {author} {\bibfnamefont {A.}~\bibnamefont
  {Altes}}, \bibinfo {author} {\bibfnamefont {R.}~\bibnamefont {Heiderhoff}},\
  and\ \bibinfo {author} {\bibfnamefont {L.~J.}\ \bibnamefont {Balk}},\
  }\bibfield  {title} {\bibinfo {title} {Quantitative thermal conductivity
  measurements with nanometre resolution},\ }\href@noop {} {\bibfield
  {journal} {\bibinfo  {journal} {J. Phys. D: Appl. Phys.}\ }\textbf {\bibinfo
  {volume} {32}},\ \bibinfo {pages} {L13} (\bibinfo {year} {1999})}\BibitemShut
  {NoStop}%
\bibitem [{\citenamefont {Sikora}\ \emph {et~al.}(2012)\citenamefont {Sikora},
  \citenamefont {Ftouni}, \citenamefont {Richard}, \citenamefont {Hebert},
  \citenamefont {Eon}, \citenamefont {Omnes},\ and\ \citenamefont
  {Bourgeois}}]{sikora2012}%
  \BibitemOpen
  \bibfield  {author} {\bibinfo {author} {\bibfnamefont {A.}~\bibnamefont
  {Sikora}}, \bibinfo {author} {\bibfnamefont {H.}~\bibnamefont {Ftouni}},
  \bibinfo {author} {\bibfnamefont {J.}~\bibnamefont {Richard}}, \bibinfo
  {author} {\bibfnamefont {C.}~\bibnamefont {Hebert}}, \bibinfo {author}
  {\bibfnamefont {D.}~\bibnamefont {Eon}}, \bibinfo {author} {\bibfnamefont
  {F.}~\bibnamefont {Omnes}},\ and\ \bibinfo {author} {\bibfnamefont
  {O.}~\bibnamefont {Bourgeois}},\ }\bibfield  {title} {\bibinfo {title}
  {Highly sensitive thermal conductivity measurements of suspended membranes
  (sin and diamond) using a 3 omega-volklein method},\ }\href
  {https://doi.org/10.1063/1.4704086} {\bibfield  {journal} {\bibinfo
  {journal} {Rev. Sci. Instrum.}\ }\textbf {\bibinfo {volume} {83}},\ \bibinfo
  {pages} {054902} (\bibinfo {year} {2012})}\BibitemShut {NoStop}%
\bibitem [{\citenamefont {Sikora}\ \emph {et~al.}(2013)\citenamefont {Sikora},
  \citenamefont {Ftouni}, \citenamefont {Richard}, \citenamefont {Hebert},
  \citenamefont {Eon}, \citenamefont {Omnes},\ and\ \citenamefont
  {Bourgeois}}]{sikora2013}%
  \BibitemOpen
  \bibfield  {author} {\bibinfo {author} {\bibfnamefont {A.}~\bibnamefont
  {Sikora}}, \bibinfo {author} {\bibfnamefont {H.}~\bibnamefont {Ftouni}},
  \bibinfo {author} {\bibfnamefont {J.}~\bibnamefont {Richard}}, \bibinfo
  {author} {\bibfnamefont {C.}~\bibnamefont {Hebert}}, \bibinfo {author}
  {\bibfnamefont {D.}~\bibnamefont {Eon}}, \bibinfo {author} {\bibfnamefont
  {F.}~\bibnamefont {Omnes}},\ and\ \bibinfo {author} {\bibfnamefont
  {O.}~\bibnamefont {Bourgeois}},\ }\bibfield  {title} {\bibinfo {title}
  {Highly sensitive thermal conductivity measurements of suspended membranes
  (sin and diamond) using a 3 omega-volklein method erratum (vol 83, 054902,
  2012)},\ }\href {https://doi.org/10.1063/1.4793652} {\bibfield  {journal}
  {\bibinfo  {journal} {Rev. Sci. Instrum.}\ }\textbf {\bibinfo {volume}
  {84}},\ \bibinfo {pages} {029901} (\bibinfo {year} {2013})}\BibitemShut
  {NoStop}%
\bibitem [{\citenamefont {Tavakoli}\ \emph {et~al.}(2022)\citenamefont
  {Tavakoli}, \citenamefont {Lulla}, \citenamefont {Puurtinen}, \citenamefont
  {Maasilta}, \citenamefont {Collin}, \citenamefont {Saminadayar},\ and\
  \citenamefont {Bourgeois}}]{tavakoli2022}%
  \BibitemOpen
  \bibfield  {author} {\bibinfo {author} {\bibfnamefont {A.}~\bibnamefont
  {Tavakoli}}, \bibinfo {author} {\bibfnamefont {K.~J.}\ \bibnamefont {Lulla}},
  \bibinfo {author} {\bibfnamefont {T.}~\bibnamefont {Puurtinen}}, \bibinfo
  {author} {\bibfnamefont {I.}~\bibnamefont {Maasilta}}, \bibinfo {author}
  {\bibfnamefont {E.}~\bibnamefont {Collin}}, \bibinfo {author} {\bibfnamefont
  {L.}~\bibnamefont {Saminadayar}},\ and\ \bibinfo {author} {\bibfnamefont
  {O.}~\bibnamefont {Bourgeois}},\ }\bibfield  {title} {\bibinfo {title}
  {Specific heat of thin phonon cavities at low temperature: Very high values
  revealed by zeptojoule calorimetry},\ }\href
  {https://doi.org/10.1103/PhysRevB.105.224313} {\bibfield  {journal} {\bibinfo
   {journal} {Phys. Rev. B}\ }\textbf {\bibinfo {volume} {105}},\ \bibinfo
  {pages} {224313} (\bibinfo {year} {2022})}\BibitemShut {NoStop}%
\bibitem [{\citenamefont {Yovanovich}\ and\ \citenamefont
  {Marotta}(2003)}]{yovanovich2003thermal}%
  \BibitemOpen
  \bibfield  {author} {\bibinfo {author} {\bibfnamefont {M.}~\bibnamefont
  {Yovanovich}}\ and\ \bibinfo {author} {\bibfnamefont {E.}~\bibnamefont
  {Marotta}},\ }\bibfield  {title} {\bibinfo {title} {Thermal spreading and
  contact resistances},\ }in\ \href@noop {} {\emph {\bibinfo {booktitle} {Heat
  Transfer Handbook}}}\ (\bibinfo {organization} {Chap. IV},\ \bibinfo {year}
  {2003})\ p.\ \bibinfo {pages} {261}\BibitemShut {NoStop}%
\bibitem [{\citenamefont {Prasher}\ and\ \citenamefont
  {Phelan}(2006)}]{prasher2006}%
  \BibitemOpen
  \bibfield  {author} {\bibinfo {author} {\bibfnamefont {R.~S.}\ \bibnamefont
  {Prasher}}\ and\ \bibinfo {author} {\bibfnamefont {P.~E.}\ \bibnamefont
  {Phelan}},\ }\bibfield  {title} {\bibinfo {title} {Microscopic and
  macroscopic thermal contact resistances of pressed mechanical contacts},\
  }\href {https://doi.org/10.1063/1.2353704} {\bibfield  {journal} {\bibinfo
  {journal} {J. Appl. Phys.}\ }\textbf {\bibinfo {volume} {100}},\ \bibinfo
  {pages} {063538} (\bibinfo {year} {2006})}\BibitemShut {NoStop}%
\bibitem [{\citenamefont {Wexler}(1966)}]{wexler1966}%
  \BibitemOpen
  \bibfield  {author} {\bibinfo {author} {\bibfnamefont {G.}~\bibnamefont
  {Wexler}},\ }\bibfield  {title} {\bibinfo {title} {Size effect and non-local
  boltzmann transport equation in orifice and disk geometry},\ }\href
  {https://doi.org/10.1088/0370-1328/89/4/316} {\bibfield  {journal} {\bibinfo
  {journal} {Proc. Phys. Soc. London}\ }\textbf {\bibinfo {volume} {89}},\
  \bibinfo {pages} {927} (\bibinfo {year} {1966})}\BibitemShut {NoStop}%
\bibitem [{\citenamefont {Nikolic}\ and\ \citenamefont
  {Allen}(1999)}]{nikolic1999}%
  \BibitemOpen
  \bibfield  {author} {\bibinfo {author} {\bibfnamefont {B.}~\bibnamefont
  {Nikolic}}\ and\ \bibinfo {author} {\bibfnamefont {P.}~\bibnamefont
  {Allen}},\ }\bibfield  {title} {\bibinfo {title} {Electron transport through
  a circular constriction},\ }\href {https://doi.org/10.1103/PhysRevB.60.3963}
  {\bibfield  {journal} {\bibinfo  {journal} {Physical Review B}\ }\textbf
  {\bibinfo {volume} {60}},\ \bibinfo {pages} {3963} (\bibinfo {year}
  {1999})}\BibitemShut {NoStop}%
\bibitem [{\citenamefont {Sharvin}(1965)}]{sharvin1965}%
  \BibitemOpen
  \bibfield  {author} {\bibinfo {author} {\bibfnamefont {Y.}~\bibnamefont
  {Sharvin}},\ }\bibfield  {title} {\bibinfo {title} {A possible method for
  studying fermi surfaces},\ }\href@noop {} {\bibfield  {journal} {\bibinfo
  {journal} {Zh. Eksp. Teor. Fiz.}\ }\textbf {\bibinfo {volume} {48}},\
  \bibinfo {pages} {984} (\bibinfo {year} {1965})}\BibitemShut {NoStop}%
\bibitem [{\citenamefont {Paterson}(2020)}]{thesejessy}%
  \BibitemOpen
  \bibfield  {author} {\bibinfo {author} {\bibfnamefont {J.}~\bibnamefont
  {Paterson}},\ }\bibfield  {title} {\bibinfo {title} {Experimental
  investigation of heat transport in nanomaterials using electro-thermal
  methods},\ }\bibfield  {journal} {\bibinfo  {journal} {Thesis Grenoble}\
  }\href@noop {} {} (\bibinfo {year} {2020})\BibitemShut {NoStop}%
\bibitem [{\citenamefont {Paterson}\ \emph {et~al.}(2020)\citenamefont
  {Paterson}, \citenamefont {Singhal}, \citenamefont {Tainoff}, \citenamefont
  {Richard},\ and\ \citenamefont {Bourgeois}}]{paterson2020}%
  \BibitemOpen
  \bibfield  {author} {\bibinfo {author} {\bibfnamefont {J.}~\bibnamefont
  {Paterson}}, \bibinfo {author} {\bibfnamefont {D.}~\bibnamefont {Singhal}},
  \bibinfo {author} {\bibfnamefont {D.}~\bibnamefont {Tainoff}}, \bibinfo
  {author} {\bibfnamefont {J.}~\bibnamefont {Richard}},\ and\ \bibinfo {author}
  {\bibfnamefont {O.}~\bibnamefont {Bourgeois}},\ }\bibfield  {title} {\bibinfo
  {title} {Thermal conductivity and thermal boundary resistance of amorphous
  al2o3 thin films on germanium and sapphire},\ }\href
  {https://doi.org/10.1063/5.0004576} {\bibfield  {journal} {\bibinfo
  {journal} {J. Appl. Phys.}\ }\textbf {\bibinfo {volume} {127}},\ \bibinfo
  {pages} {245105} (\bibinfo {year} {2020})}\BibitemShut {NoStop}%
\bibitem [{\citenamefont {Hoogeboom-Pot}\ \emph {et~al.}(2015)\citenamefont
  {Hoogeboom-Pot}, \citenamefont {Hernandez-Charpak}, \citenamefont {Gu},
  \citenamefont {Frazer}, \citenamefont {Anderson}, \citenamefont {Chao},
  \citenamefont {Falcone}, \citenamefont {Yang}, \citenamefont {Murnane},
  \citenamefont {Kapteyn},\ and\ \citenamefont {Nardi}}]{hoogeboom2015}%
  \BibitemOpen
  \bibfield  {author} {\bibinfo {author} {\bibfnamefont {K.~M.}\ \bibnamefont
  {Hoogeboom-Pot}}, \bibinfo {author} {\bibfnamefont {J.~N.}\ \bibnamefont
  {Hernandez-Charpak}}, \bibinfo {author} {\bibfnamefont {X.}~\bibnamefont
  {Gu}}, \bibinfo {author} {\bibfnamefont {T.~D.}\ \bibnamefont {Frazer}},
  \bibinfo {author} {\bibfnamefont {E.~H.}\ \bibnamefont {Anderson}}, \bibinfo
  {author} {\bibfnamefont {W.}~\bibnamefont {Chao}}, \bibinfo {author}
  {\bibfnamefont {R.~W.}\ \bibnamefont {Falcone}}, \bibinfo {author}
  {\bibfnamefont {R.}~\bibnamefont {Yang}}, \bibinfo {author} {\bibfnamefont
  {M.~M.}\ \bibnamefont {Murnane}}, \bibinfo {author} {\bibfnamefont {H.~C.}\
  \bibnamefont {Kapteyn}},\ and\ \bibinfo {author} {\bibfnamefont
  {D.}~\bibnamefont {Nardi}},\ }\bibfield  {title} {\bibinfo {title} {A new
  regime of nanoscale thermal transport: Collective diffusion increases
  dissipation efficiency},\ }\href {https://doi.org/10.1073/pnas.1503449112}
  {\bibfield  {journal} {\bibinfo  {journal} {Proc. Nat. Acad. Sci.}\ }\textbf
  {\bibinfo {volume} {112}},\ \bibinfo {pages} {4846} (\bibinfo {year}
  {2015})}\BibitemShut {NoStop}%
\bibitem [{\citenamefont {Alikin}\ \emph {et~al.}(2023)\citenamefont {Alikin},
  \citenamefont {Zakharchuk}, \citenamefont {Xie}, \citenamefont {Romanyuk},
  \citenamefont {Pereira}, \citenamefont {Arias-Serrano}, \citenamefont
  {Weidenkaff}, \citenamefont {Kholkin}, \citenamefont {Kovalevsky},\ and\
  \citenamefont {Tselev}}]{alikin2023}%
  \BibitemOpen
  \bibfield  {author} {\bibinfo {author} {\bibfnamefont {D.}~\bibnamefont
  {Alikin}}, \bibinfo {author} {\bibfnamefont {K.}~\bibnamefont {Zakharchuk}},
  \bibinfo {author} {\bibfnamefont {W.}~\bibnamefont {Xie}}, \bibinfo {author}
  {\bibfnamefont {K.}~\bibnamefont {Romanyuk}}, \bibinfo {author}
  {\bibfnamefont {M.~J.}\ \bibnamefont {Pereira}}, \bibinfo {author}
  {\bibfnamefont {B.~I.}\ \bibnamefont {Arias-Serrano}}, \bibinfo {author}
  {\bibfnamefont {A.}~\bibnamefont {Weidenkaff}}, \bibinfo {author}
  {\bibfnamefont {A.}~\bibnamefont {Kholkin}}, \bibinfo {author} {\bibfnamefont
  {A.~V.}\ \bibnamefont {Kovalevsky}},\ and\ \bibinfo {author} {\bibfnamefont
  {A.}~\bibnamefont {Tselev}},\ }\bibfield  {title} {\bibinfo {title}
  {Quantitative characterization of local thermal properties in thermoelectric
  ceramics using ``jumping-mode{''} scanning thermal microscopy},\ }\href
  {https://doi.org/10.1002/smtd.202201516} {\bibfield  {journal} {\bibinfo
  {journal} {Small Methods}\ }\textbf {\bibinfo {volume} {7}},\ \bibinfo
  {pages} {14} (\bibinfo {year} {2023})}\BibitemShut {NoStop}%
\bibitem [{\citenamefont {Gonzalez-Munoz}\ \emph {et~al.}(2023)\citenamefont
  {Gonzalez-Munoz}, \citenamefont {Agarwal}, \citenamefont {Castanon},
  \citenamefont {Kudrynskyi}, \citenamefont {Kovalyuk}, \citenamefont {Spiece},
  \citenamefont {Kazakova}, \citenamefont {Patane},\ and\ \citenamefont
  {Kolosov}}]{gonzalez2023}%
  \BibitemOpen
  \bibfield  {author} {\bibinfo {author} {\bibfnamefont {S.}~\bibnamefont
  {Gonzalez-Munoz}}, \bibinfo {author} {\bibfnamefont {K.}~\bibnamefont
  {Agarwal}}, \bibinfo {author} {\bibfnamefont {E.~G.}\ \bibnamefont
  {Castanon}}, \bibinfo {author} {\bibfnamefont {Z.~R.}\ \bibnamefont
  {Kudrynskyi}}, \bibinfo {author} {\bibfnamefont {Z.~D.}\ \bibnamefont
  {Kovalyuk}}, \bibinfo {author} {\bibfnamefont {J.}~\bibnamefont {Spiece}},
  \bibinfo {author} {\bibfnamefont {O.}~\bibnamefont {Kazakova}}, \bibinfo
  {author} {\bibfnamefont {A.}~\bibnamefont {Patane}},\ and\ \bibinfo {author}
  {\bibfnamefont {O.~V.}\ \bibnamefont {Kolosov}},\ }\bibfield  {title}
  {\bibinfo {title} {Direct measurements of anisotropic thermal transport in
  gamma-inse nanolayers via cross-sectional scanning thermal microscopy},\
  }\href {https://doi.org/10.1002/admi.202300081} {\bibfield  {journal}
  {\bibinfo  {journal} {Adv. Mat. Interfaces}\ }\textbf {\bibinfo {volume}
  {10}},\ \bibinfo {pages} {2300081} (\bibinfo {year} {2023})}\BibitemShut
  {NoStop}%
\bibitem [{\citenamefont {Umatova}\ \emph {et~al.}(2019)\citenamefont
  {Umatova}, \citenamefont {Zhang}, \citenamefont {Rajkumar}, \citenamefont
  {Dobson},\ and\ \citenamefont {Weaver}}]{umatova2019}%
  \BibitemOpen
  \bibfield  {author} {\bibinfo {author} {\bibfnamefont {Z.}~\bibnamefont
  {Umatova}}, \bibinfo {author} {\bibfnamefont {Y.}~\bibnamefont {Zhang}},
  \bibinfo {author} {\bibfnamefont {R.}~\bibnamefont {Rajkumar}}, \bibinfo
  {author} {\bibfnamefont {P.~S.}\ \bibnamefont {Dobson}},\ and\ \bibinfo
  {author} {\bibfnamefont {J.~M.~R.}\ \bibnamefont {Weaver}},\ }\bibfield
  {title} {\bibinfo {title} {Quantification of atomic force microscopy tip and
  sample thermal contact},\ }\href {https://doi.org/10.1063/1.5097862}
  {\bibfield  {journal} {\bibinfo  {journal} {Rev. Sci. Instrum.}\ }\textbf
  {\bibinfo {volume} {90}},\ \bibinfo {pages} {095003} (\bibinfo {year}
  {2019})}\BibitemShut {NoStop}%
\bibitem [{\citenamefont {Zhang}\ \emph {et~al.}(2012)\citenamefont {Zhang},
  \citenamefont {Dobson},\ and\ \citenamefont {Weaver}}]{zhang2012high}%
  \BibitemOpen
  \bibfield  {author} {\bibinfo {author} {\bibfnamefont {Y.}~\bibnamefont
  {Zhang}}, \bibinfo {author} {\bibfnamefont {P.~S.}\ \bibnamefont {Dobson}},\
  and\ \bibinfo {author} {\bibfnamefont {J.~M.~R.}\ \bibnamefont {Weaver}},\
  }\bibfield  {title} {\bibinfo {title} {High temperature imaging using a
  thermally compensated cantilever resistive probe for scanning thermal
  microscopy},\ }\href {https://doi.org/10.1116/1.3664328} {\bibfield
  {journal} {\bibinfo  {journal} {J. Vac. Sci. Techno. B}\ }\textbf {\bibinfo
  {volume} {30}},\ \bibinfo {pages} {010601} (\bibinfo {year}
  {2012})}\BibitemShut {NoStop}%
\end{thebibliography}%
	
\end{document}